\pdfoutput=1
\documentclass[aps,prx,twocolumn,longbibliography,superscriptaddress,balancelastpage]{revtex4-2}
\usepackage{amsmath}
\allowdisplaybreaks
\DeclareMathOperator{\sgn}{sgn}
\usepackage{bm}
\usepackage{bbold}
\usepackage{amssymb}
\usepackage{graphicx}
\usepackage[ hidelinks = true, colorlinks = true, citecolor = blue, urlcolor = blue, linkcolor = blue]{hyperref}
\usepackage{verbatim}
\usepackage{color}
\usepackage[capitalise]{cleveref}
\usepackage[normalem]{ulem}
\usepackage{dsfont}
\usepackage{multirow}

\allowdisplaybreaks[3]

\usepackage{accents}
\newlength{\dhatheight}
\newcommand{\doublehat}[1]{%
    \settoheight{\dhatheight}{\ensuremath{\hat{#1}}}%
    \addtolength{\dhatheight}{-0.35ex}%
    \hat{\vphantom{\rule{1pt}{\dhatheight}}%
    \smash{\hat{#1}}}}

\begin{document}

\title{Metastability and quantum coherence-assisted sensing \\in interacting parallel quantum dots }

\author{Stephanie Matern}
\affiliation{NanoLund and Solid State Physics,  Lund University,  Box 118,  22100 Lund, Sweden}
\author{Katarzyna Macieszczak}
\affiliation{Department of Physics, University of Warwick, Coventry CV4 7AL, United Kingdom}
\affiliation{TCM Group, Cavendish Laboratory, University of Cambridge, J. J. Thomson Ave., Cambridge CB3 0HE, United Kingdom}
\author{Simon Wozny}
\affiliation{NanoLund and Solid State Physics,  Lund University,  Box 118,  22100 Lund, Sweden}
\author{Martin Leijnse}
\affiliation{NanoLund and Solid State Physics,  Lund University,  Box 118,  22100 Lund, Sweden}

\date{\today}

\begin{abstract}
We study the transient dynamics subject to quantum coherence effects of two interacting parallel quantum dots weakly coupled to macroscopic leads.  The stationary particle current of this quantum system is sensitive to perturbations much smaller than any other energy scale, specifically compared to the system-lead coupling and the temperature. We show that this is due to the presence of a  parity-like symmetry in the dynamics, as a consequence of which, two distinct stationary states arise. In the presence of small perturbations breaking this symmetry, the system exhibits metastability with two metastable phases that can be approximated by a combination of states corresponding to stationary states in the unperturbed limit. Furthermore,  the long-time dynamics can be described as classical dynamics between those phases, leading to a unique stationary state. In particular, the competition of those two metastable phases explains the sensitive behavior of the stationary current towards small perturbations.
We show that this behavior bears the potential of utilizing the parallel dots as a charge sensor which makes use of quantum coherence effects to achieve a signal to noise ratio that is not limited by the temperature.  As a consequence, the parallel dots outperform an analogous single-dot charge sensor for a wide range of temperatures.
\end{abstract}

\maketitle

%%%%%%%%%%%%%%%%%%%%%%%%%%%%%%%%%%%%%%%%%%%%%%%%%%%%%%%%%%%%%%%%%%%%%%%
%                  SEC: INTRODUCTION
%%%%%%%%%%%%%%%%%%%%%%%%%%%%%%%%%%%%%%%%%%%%%%%%%%%%%%%%%%%%%%%%%%%%%%%
%

\section{Introduction}

The coherent control of electronic quantum devices is a challenging necessity for many novel device concepts \cite{Bauerle2018}. 
Some device concepts are based on quantum coherent non-equilibrium charge transport, while in other cases measurements of charge transport can be used to gain information about the quantum system.
In many cases, the stationary properties are of interest, but with increasing control over quantum systems understanding the full transient behavior is of importance. 
For any such application  the dynamics beyond the stationary state becomes relevant and attracted much attention in recent years \cite{Ridley2022,Weymann2021,Contreras2012,Schulenborg2014,Cheng2018,Taranko2019}.  Generally, the relaxation of a quantum system can be complicated and it can experience a quasi-stationary state before relaxing into its true stationary state. 
This phenomenon of metastability \cite{Bovier2002,Macieszczak2016,Rose2016,Macieszczak2021} occurs when the timescales dictating the system's dynamics are well separated.  Importantly,  metastability always occurs for systems in the proximity to multi-stable points where the dynamics features multiple stationary states, which may arise, e.g., due to a symmetry of the dynamical equations.
It can also be observed in constrained systems such as quantum spin glasses \cite{Cugliandolo1999, Olmos2012}, quantum gases \cite{Letscher2017,Olmos2014} and superconducting nanojunctions \cite{Souto2017}.

Another ingredient to understand a quantum system's behavior lies with quantum interference and coherence effects and how they affect the system's properties and dynamics.  
Different types of quantum dot systems constitute particularly well-controlled and versatile platforms to study and use various aspects of interference effects \cite{deArquer2021,Goldstein2007,Donarini2019,Karrasch2007a}.
It has been suggested that they can, e.g., reduce power fluctuations and rectify heat transport \cite{Ptaszynski2018,Vannucci2015}, or affect thermoelectric properties \cite{Miao2018} and electronic transport properties \cite{Lambert2015} in molecules. One can also exploit the quantum interference to construct a transistor \cite{Stafford2007}.
Particularly, interference effects in the transport properties of parallel double quantum dots have been a longstanding topic of interest. On the one hand, in the strong coupling regime, strongly-correlated physics 
dominates the transport exhibiting the Kondo effect and leading to defined signatures in the conductance as well as population switching \cite{Boese2001,Meden2006,Kashcheyevs2007,Kashcheyevs2009,Lee2007}.

On the other hand,  the coherence effects present in parallel double quantum dots, understood as a superposition state of a single fermion on either the upper or lower dot,  have been shown to play a role for quantum thermodynamics \cite{Cuetara2016}, and impact the thermoelectric current \cite{Sierra2016}, thermal conductance \cite{Zhang2021} and electric transport  \cite{Schaller2009,Li2019} in the weak coupling limit.

\begin{figure}[t!]
	\centering
	\includegraphics[width = 0.4\textwidth]{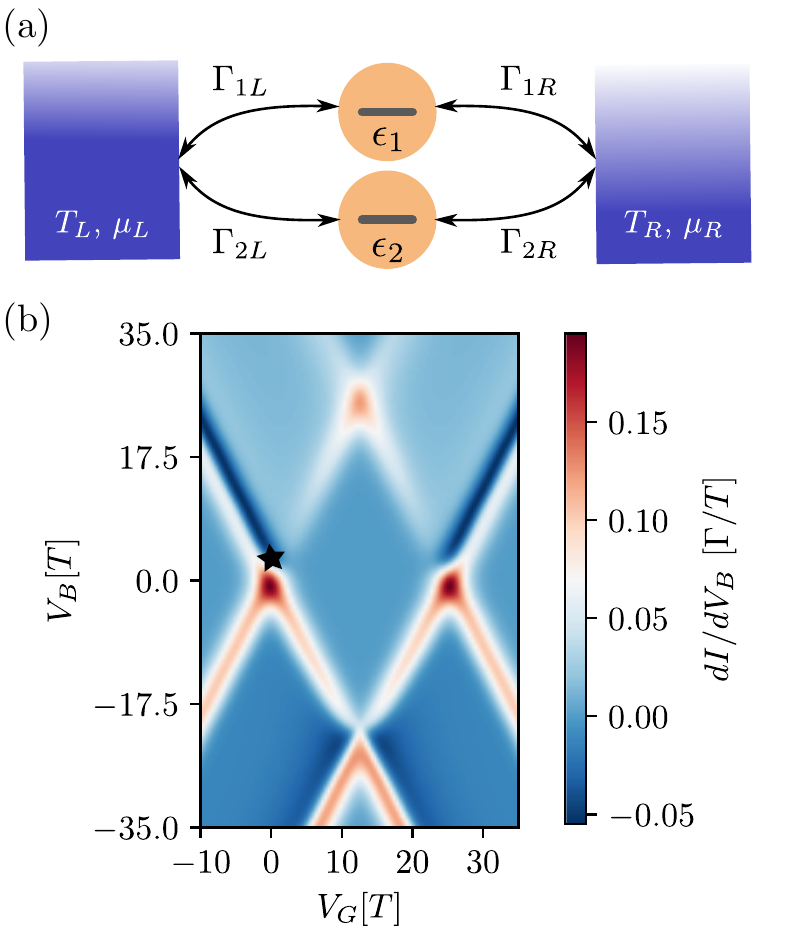}
	\caption{(a) The parallel dots with energies $\epsilon_i$ are coupled to  macroscopic leads with the tunneling rates $\Gamma_{js}$ between  dot $j$ and lead $s$, where $j=1,2$ and $s=L,R$. A bias is applied across the left and right leads, with temperatures $T_L$ and  $T_R$ and chemical potentials $\mu_L$ and  $\mu_R$, respectively.  (b)  $d I/d V_B$ as a function of the bias voltage  $V_B$ and the gate voltage $V_G$.  In Figs.~\ref{fig:model}-\ref{fig:sensor}, and~\ref{fig:stab_app}, we consider perturbations away from the balanced setup, that is Eqs.~\eqref{eq:per_epsilon} and~\eqref{eq:per_Gamma}. Here, $\delta \epsilon = \delta \Gamma = 0.04 \Gamma$ and $U = 250 \Gamma, T_L = T_R = 10 \Gamma$. The star marks the voltage parameters $V_G = 0 \Gamma$ and $V_B = 30 \Gamma $ used in other figures.}
	\label{fig:model}
\end{figure}

 In this paper we use a quantum master equation approach to study the non-equilibrium transport properties of such a parallel dots system, sketched in \cref{fig:model}(a). The stationary state of this quantum system has been shown, in certain parameter regimes, to be sensitive to small perturbations in the system-lead coupling and the detuning of the dot energies \cite{Li2019,Schaller2009}.  Remarkably,  due to quantum coherence effects, changes that are much smaller than all other energy scales in the system can lead to large changes in the stationary particle current \cite{Li2019}.

 We show that the sensitive response results from the presence of two stationary states when both the dot energies and their tunnel couplings to the leads are identical.  Perturbations breaking the corresponding  symmetry introduce a large  timescale which is well separated from any other timescale in the system's dynamics.  For intermediate times,  metastability occurs, and the state of the system is well approximated by a probabilistic mixture of two metastable phases \cite{Macieszczak2016,Rose2016,Macieszczak2021}. 
In the long time limit,  those probabilities evolve according to classical dynamics dominated by the emergent timescale due to the broken symmetry.  As the long-time dynamics depend not only on the size but also on the structure of perturbations breaking the symmetry,  the stationary state does as well, and the stationary current displays large changes resulting from small parameter changes. In particular, we focus on the regime of large Coulomb interaction where one of the metastable phases features suppressed particle current through the system, while the other phase supports a much larger current, so that the stationary current varies from suppressed to larger values.
 
Furthermore, we investigate how the sensitive behavior of the current can be used to enable the system to act as a charge sensor. 
To quantify the accuracy of the sensor we also need the current noise, which we calculate based on counting statistics \cite{Emary2009,Kirsanskas2017}. One problem for sensing applications is that metastability typically results in large current noise. Another problem is the long relaxation times associated with metastability. 
Nonetheless, we show that there is a large parameter regime where the parallel dots by far outperform a single dot used as a charge sensor.

The paper is organized as follows.  
After introducing the model in \cref{sec:model}, we discuss the Lindblad dynamics of the parallel dots in \cref{sec:Lindblad_dynamics}. In \cref{sec:transient} we discuss the transient dynamics. 
  This includes a discussion of the unperturbed and perturbed dynamics in the context of symmetry breaking and the resulting metastability and the long time dynamics towards a unique stationary state.  Finally, in \cref{sec:sensing} we investigate  how the parallel dots could be used as a charge sensor.

%%%%%%%%%%%%%%%%%%%%%%%%%%%%%%%%%%%%%%%%%%%%%%%%%%%%%%%%%%%%%%%%%%%%%%%

%%%%%%%%%%%%%%%%%%%%%%%%%%%%%%%%%%%%%%%%%%%%%%%%%%%%%%%%%%%%%%%%%%%%%%%
%                  SEC: model 
%%%%%%%%%%%%%%%%%%%%%%%%%%%%%%%%%%%%%%%%%%%%%%%%%%%%%%%%%%%%%%%%%%%%%%%
%

\section{Model} \label{sec:model}

The setup under consideration consists of two interacting parallel quantum dots weakly coupled to macroscopic leads, see \cref{fig:model}(a). 
To simplify the analytic treatment we consider spinless electrons in the following, but we have verified by direct comparison that the qualitative physics and results remain the same for spin-degenerate dot orbitals (our spinless model can be realized in a system where the Zeeman energy is larger than the applied bias voltage).

The Hamiltonian of the entire setup splits into three parts,
\begin{align}
	H = H_\text{PD} + H_L + H_T, 
	\label{eq:H}
\end{align}
for the parallel dots, the leads and the interaction between the subsystems.  The dot Hamiltonian is given by 
\begin{align}
	H_\text{PD} = \sum_{j=1,2} \epsilon_j d^\dagger_j d_j + U d^\dagger_1 d_1 d^\dagger_2 d_2
	\label{eq:HQD}
\end{align}
with the fermionic creation and annihilation operators,  $d^\dagger_j$ and $d_j$, where $j = 1,2$ is the dot label, the energy levels of the dots are $\epsilon_j$ and $U$ denotes the Coulomb interaction; we set $\hbar=1$ throughout the paper.

The leads are described by non-interacting fermions,
\begin{align}	\label{eq:HL}
	H_L =\sum_{ s = L,R}\sum_{k} \omega_{sk} c^\dagger_{sk} c_{sk}.
\end{align}
The operators $c^\dagger_{ks }, \ c_{sk}$ create or annihilate an electron in the left ($s=L$) or right ($s=R$) lead at momentum $k$, with their corresponding energy dispersion given by $\omega_{sk}$.  
Finally,  the tunneling between the parallel dots and the leads is governed by
\begin{align}
	H_T = \sum_{j = 1,2}\sum_{ s = L,R}\sum_{k} \left( t_{jsk} c^\dagger_{sk} d_j +  t^*_{jsk} d^\dagger_j c_{sk}\right),
	\label{eq:HT}
\end{align}
with the tunneling amplitude $t_{jsk}$.
We note that $H_T$ assumes both dots to couple to the same lead channel. This is a crucial assumption for the results of this work, which will hold for one-dimensional leads, but is also justified when the dots are close on a length scale set by the Fermi wavelength in the leads. 
We furthermore take $t_{jsk}$ to be real and positive (we can always make this choice if relative phases between $t_{1sk}$ and $t_{2sk}$ are independent of $k$ and $s$).  If a magnetic field is present, or the leads are ferromagnetic or superconducting, one might need to consider complex $t_{jsk}$, which possibly leads to additional phenomena, such as phase lapses in the conductance \cite{Meden2006,Karrasch2007b,Kashcheyevs2007,Golosov2007}.

The electrons in the leads are assumed to be described by the grand canonical ensemble with the chemical potentials $\mu_s$ and temperatures  $T_s$; we set Boltzmann's constant $k_B=1$ and the elementary charge $e=1$. 
The chemical potentials can be controlled by the bias voltage $V_B=2\mu_L=-2\mu_R$, while the gate allows control of the dot energy levels,  $V_G = -(\epsilon_1 + \epsilon_2)/2$.

%%%%%%%%%%%%%%%%%%%%%%%%%%%%%%%%%%%%%%%%%%%%%%%%%%%%%%%%%%%%%%%%%%%%%%%

%%%%%%%%%%%%%%%%%%%%%%%%%%%%%%%%%%%%%%%%%%%%%%%%%%%%%%%%%%%%%%%%%%%%%%%
%                  SEC: Lindblad dynamics
%%%%%%%%%%%%%%%%%%%%%%%%%%%%%%%%%%%%%%%%%%%%%%%%%%%%%%%%%%%%%%%%%%%%%%%
%

\section{Lindblad dynamics of ~parallel~dots} \label{sec:Lindblad_dynamics}

\subsection{Master equation}

In this work, we focus on the limit of weak tunnel coupling. In this case, the dynamics of the reduced density matrix $\rho_\text{PD}(t)$ of the two quantum dots can be  well approximated by a  quantum master equation of Gorini-Kossakowski-Lindblad-Sudarshan (GKLS) form  \cite{Gorini1976,Lindblad1976}. 

We assume that tunneling amplitudes are independent of  the momentum  $t_{jsk}= t_{js}$,  and that the density of states is constant, $\nu_{sk} = \nu$.  The tunneling amplitudes $t_{js}$ define the tunneling rates as $\Gamma_{js} =2 \pi \nu\lvert t_{js}   \lvert^2$.  The weak coupling limit where our master equation is valid is defined by $\Gamma_{js}\ll T_s$.

 Following  \cite{Kirsanskas2018,Nathan2020,Ptaszy2019},   we consider terms quadratic in tunneling amplitudes to arrive at the master equation beyond the secular approximation,
\begin{equation}
\begin{aligned}
	&\frac{d}{d t} \rho_\text{PD}(t) = - i \left[H_\text{eff}, \rho_\text{PD}(t)\right]\\
	& + \sum_{\substack{\alpha = +,-\\ s = L,R}} \left[ J_{\alpha s} \rho_\text{PD}(t) J_{\alpha s}^\dagger - \frac{1}{2}\left\{ \rho_\text{PD}(t), J_{\alpha s}^\dagger J_{\alpha s}\right\}\right],
	\label{eq:evolution}
\end{aligned}
\end{equation}
where  $[\cdot, \cdot ]$ and $\{\cdot, \cdot \}$ stand for the commutator and anti-commutator, respectively. Here, the effective Hamiltonian $H_\text{eff} = H_\text{PD} + H_\text{LS}$  includes a Lamb shift $H_\text{LS}$,  which renormalizes the energies of the parallel dots and can be identified as an effective tunnel splitting \cite{Boese2001}.  We exclude a direct interdot tunnel coupling of the form $ \Omega_{12} d_1^\dagger d_2 + \text{h.c.}$ in $H_\text{PD}$ of \cref{eq:HQD}, but such a term can easily be added to $H_\text{PD}$ and the qualitative physics described in the following does not change for  $\Omega_{12}\ll T_s$.

The jump operators  $J_{\alpha s}$ describe the exchange of an electron between the parallel dots and lead $s$ with $\alpha~=~+$ representing an electron entering the dots and $\alpha~=~-$  an electron leaving.   The closed form expressions for the Lamb shift and  the  jump operators are derived in \cref{app:Lindblad}.

%%%%%%%%%%%%%%%%%%%%%%%%%%%%%%%%%%%%%%%%%%%%%%%%%%%%%%%%%%%%%%%%%%%%%%%
%                  subSEC: Metastability
%%%%%%%%%%%%%%%%%%%%%%%%%%%%%%%%%%%%%%%%%%%%%%%%%%%%%%%%%%%%%%%%%%%%%%%
%

\subsection{Spectral decomposition and metastability}

The equation of motion in \cref{eq:evolution} can be recast as
\begin{align}
	\frac{d}{d t}  \rho_\text{PD}(t)  = {\mathcal{L}}  \rho_\text{PD}(t),
	\label{eq:evolution2}
\end{align}
where $\mathcal{L}$ is the  Liouvillian.
%so-called \emph{master operator}. 
Therefore, the full time evolution, which is a completely positive, trace-preserving map, can be formally solved as
\begin{align}
	 \rho_\text{PD}(t)  = e^{t\mathcal{L}}  \rho_\text{PD}(0).
	\label{eq:evolution2}
\end{align}
It follows that the evolution can be decomposed in terms of the Liouville operator spectrum, that is, its eigenvalues  $\lambda_i$ and left and right eigenmatrices,  $L_i $ and $R_i$, as
\begin{align}
	\rho_\text{PD}(t)  = \rho_\text{PD}^\text{ss} + \sum_{i\geq 2} e^{\lambda_i t} c_i R_i,
	\label{eq:evolution3}
\end{align}
with the coefficients $c_i = \text{Tr}[ L_i \rho_\text{PD}(0)]$ and the left and right eigenmatrices normalized so that $\text{Tr}(L_i R_j) =\delta_{ij}$.  Here, the eigenvalues are ordered with a decreasing real part, so that $\lambda_1 = 0$ corresponds to a stationary state $R_1=\rho_\text{PD}^\text{ss}$, while $L_1=\mathds{1}$, where $\mathds{1}$ is the identity operator on the dots. When the stationary state is unique, the sum runs over the decay modes with $\mathrm{Re}(\lambda_i)<0$, so that 	$\lim_{t\rightarrow\infty}\rho_\text{PD}(t) =\rho_\text{PD}^\text{ss} $ for any initial state.

If there exists a large difference in the real parts of the second and third eigenvalues, $-\lambda_2 \ll -\text{Re}(\lambda_3)$, metastability arises followed by long-time dynamics dominated by the single low-lying eigenmode \cite{Macieszczak2016,Rose2016}. Here, $\lambda_2$ is necessarily real as  $\mathcal{L}$ preserves the Hermiticity of $\rho_\text{PD}(t) $ and therefore complex eigenvalues need to appear as complex conjugate pairs. Indeed,  for times $t$ such that $ - \text{Re}(\lambda_3) t  \gg 1$, the state of the system can be approximated  as   
\begin{align}
	\rho_\text{PD}(t)  \approx  \rho_\text{PD}^\text{ss} + e^{\lambda_2 t} c_2 R_2 
	\label{eq:evolution_meta}
\end{align} 
For times $t$ such that  $-\lambda_2 t\ll 1$, the system is metastable, with its state approximated by a linear combination of the stationary state and the low-lying eigenmode, $\rho_\text{PD}(t)~\approx  \rho_\text{PD}^\text{ss} +  c_2 R_2 $,
where $c_2$ carries the information about the initial condition. For longer  times, the decay of the low-lying mode in \cref{eq:evolution_meta} can no longer be neglected. In particular, when  $  - \lambda_2 t \gg 1$, the system state approaches its asymptotic limit and is well approximated by the stationary state,  $ \rho_\text{PD}^\text{ss}$, independently of the initial condition.
 Thus,  $-1/\text{Re}(\lambda_3)$ and $ -1/\lambda_2$ can be considered as the timescales of the initial and final relaxation, respectively. 
 
 In this work, we show that for the dynamics in \cref{eq:evolution}, such metastability emerges as a consequence of breaking a parity-like symmetry originating from the Hamiltonian in \cref{eq:H}. 
For the case of a single low-lying eigenvalue, metastable states correspond to probabilistic mixtures of two metastable phases and can be investigated numerically \cite{Macieszczak2016,Rose2016}, see also \cref{app:pt_meta}. Here, we uncover the metastable phases, together with the unique stationary state analytically by means of non-Hermitian perturbation theory \cite{Kato}.

%%%%%%%%%%%%%%%%%%%%%%%%%%%%%%%%%%%%%%%%%%%%%%%%%%%%%%%%%%%%%%%%%%%%%%%
%                  subSEC: particle currents
%%%%%%%%%%%%%%%%%%%%%%%%%%%%%%%%%%%%%%%%%%%%%%%%%%%%%%%%%%%%%%%%%%%%%%%

\subsection{Dynamics of particle currents}

The average particle current leaving lead $s$ at time $t$ is given by $I_s(t)=-i\langle\left[H, N_s\right]\rangle $, with the electron number operator in the lead $N_s = \sum_k{c^\dagger_{sk}} c_{sk}$. 
Within the Lindblad dynamics, the current $I_s(t)$  is given in terms of the jump operator by \cite{Macieszczak2016b, Kirsanskas2017}
\begin{align}
	I_s(t) = \sum_{\alpha = +,-} \alpha \,\text{Tr}\left[J_{\alpha s}^\dagger J_{\alpha s} \rho_\text{PD}(t)\right],
    \label{eq:current}
\end{align}
 Asymptotically, the currents equilibrate, $I_L+I_R=0$, with $I_s=\lim_{t\rightarrow\infty} I_s(t)$ denoting the stationary current leaving lead $s$. 
 In the remainder of the paper, we will therefore study the current $I_L(t)$ leaving the left lead and we will omit the lead index. Beyond its average the current dynamics can be investigated in terms of full counting statistics \cite{Schaller,Flindt2008,Emary2009} used in \cref{sec:sensing},  see also \cref{app:counting_statistics}.

\Cref{fig:model}(b) shows the differential conductance $d I/ d V_B$ as a function of $V_G$ and $V_B$.  It has a much richer structure than the results for a double dot system with a density matrix assumed to be diagonal in the eigenbasis of $H_\text{PD}$  and thus evolving according to a Pauli  rate equation, rather than by \cref{eq:evolution}; see~\cref{app:Pauli_stability}.  This is due to coherences between singly occupied states playing a non-negligible role  in the properties of both the dynamics and the stationary state, especially when model parameters are chosen in the proximity to those for which strong symmetries are present. But we also clarify in which limits the Pauli rate equation reproduces the dynamics.

The stationary limit of the transport properties of the parallel dots system has been studied before \cite{Li2019, Schaller2009} and it has been reported that the stationary current may be highly sensitive to perturbations in the tunneling rates and to detuning of the dot energies \cite{Li2019}, see also \cref{fig:sensor}(a).
In this work, we explain this phenomenon in relation to symmetry breaking and demonstrate that the sensitivity of the stationary current can in fact be arbitrarily large.  We then verify its usefulness for sensing applications.

%%%%%%%%%%%%%%%%%%%%%%%%%%%%%%%%%%%%%%%%%%%%%%%%%%%%%%%%%%%%%%%%%%%%%%%

 %%%%%%%%%%%%%%%%%%%%%%%%%%%%%%%%%%%%%%%%%%%%%%%%%%%%%%%%%%%%%%%%%%%%%%%
%                  SEC: Symmetry and Dynamics
%%%%%%%%%%%%%%%%%%%%%%%%%%%%%%%%%%%%%%%%%%%%%%%%%%%%%%%%%%%%%%%%%%%%%%%

\section{Symmetries and Dynamics} \label{sec:transient}

We now discuss symmetries of the Hamiltonian dynamics for the total setup and the resulting properties of the Lindblad dynamics of the dots. In particular, we show how two distinct stationary states occur as a result of a swap symmetry between the dots present in the Hamiltonian. We then analyze the metastability arising by perturbatively breaking this symmetry, the long-time dynamics that follows, and the resulting unique stationary state and the associated current. Crucially, the structure of  perturbations affects the stationary state already in the leading order.

\subsection{Weak and strong symmetries}\label{sec:weak}

The Hamiltonian $H$ in \cref{eq:H} conserves the total number of electrons,  i.e.,  $[H, N_\text{PD} + \sum_{s=L,R} N_s]=0$, with $N_\text{PD}$ and $N_s$ the number operator for the parallel dots and for the lead $s$, respectively.
Since the leads feature no coherences between states with different numbers of electrons $N_s$,
any $\rho_\text{PD}$ that is diagonal in charge at the intial time will remains so at all times. 

In the approximation of the Lindblad dynamics this symmetry is inherited as a weak symmetry of the Liouvillian with respect to $N_\text{PD}$, that is, $[\mathcal{L},\mathcal{N}_\text{PD}] = 0$, where $\mathcal{N}_\text{PD} \rho_\text{PD} = [N_\text{PD}, \rho_\text{PD} ]$ \cite{Buca2012,Albert2014}, see also  \cref{app:Lindblad}. 
Corresponding density matrices feature at most six non-zero entries in the  basis of $\lvert 00 \rangle,  \lvert 10\rangle=d_1^\dagger \lvert 00 \rangle,  \lvert 01\rangle=d_2^\dagger \lvert 00 \rangle  $,  and $\lvert 11 \rangle=d_1^\dagger d_2^\dagger \lvert 00\rangle $, which is the eigenbasis of $H_\text{PD}$ in \cref{eq:HQD} and will be referred to as the \emph{local basis}. In that case, at most six modes contribute in \cref{eq:evolution3}.\\

We now discuss symmetries of the Hamiltonian and Lindblad dynamics originating from degenerate energies of the dots and their identical couplings to the two leads.
Let us consider tunneling amplitudes such that
\begin{align}\label{eq:sym_par0}
	t_{jsk}=t_{k},
\end{align}
and consider degenerate dot energies, 
 \begin{align}\label{eq:sym_par1}
 	\epsilon_j&=\epsilon.
 \end{align}
In the local basis, the Hamiltonian in \cref{eq:H}, which we denote by $H^{(0)}$ to indicate that the above conditions are fulfilled, remains the same when swapping the dot labels, up to the change of the sign for the doubly occupied state for the Coulomb interaction term. 
Thus, it  is left invariant by the swap operator $S$, $[H^{(0)},S]=0$, which exchanges the fermionic excitations between the dots,
\begin{align}
	S =\lvert00\rangle\!\langle00\rvert + \lvert 10\rangle\!\langle01 \rvert + \lvert 01\rangle\!\langle10 \rvert - \lvert11\rangle\!\langle11\rvert.
	\label{eq:swap}
\end{align}
We have that $S^2=\mathds{1}$, so $S$ is a parity operator. Indeed, the basis $\left\{\vert 00\rangle, \lvert + \rangle, \lvert - \rangle, \lvert 11\rangle\right\},$ where $\lvert \pm \rangle = \left(\lvert 10 \rangle \pm \lvert 01 \rangle\right)/\sqrt{2}$, we have $S=|00\rangle\!\langle 00|+|+\rangle\!\langle +|-|-\rangle\!\langle -|-|11\rangle\!\langle 11|$. We refer to $\lvert + \rangle$ and $\lvert - \rangle$ as the bonding and anti-bonding states, even though they remain degenerate here because of the absence of hybridization between the dots, and to the basis of $|00\rangle$, $|+\rangle$, $|-\rangle$, and $|11\rangle$ as \emph{bonding/anti-bonding basis}.

For $t_{jsk}=t_{js}$ assumed in the derivation of the Lindblad dynamics in \cref{eq:evolution}, the condition in \cref{eq:sym_par0} can be expressed as the tunneling rates being equal \cite{Li2019}
\begin{equation}\label{eq:sym_par2}
 \Gamma_{js}=\Gamma.
\end{equation}
The Liouvillian inherits the symmetry of the Hamiltonian as a strong symmetry \cite{Buca2012, Albert2014} with respect to $S$,  i.e., the symmetries of the effective Hamiltonian, $[S,H_\text{eff}^{(0)} ]=0$, and, in contrast to a weak symmetry, additionally the jump operators, $[S,J_{\alpha s}^{(0)} ]=0$. Here, we used the index $^{(0)}$ to indicate that the conditions in \cref{eq:sym_par1} and~\eqref{eq:sym_par2} are fulfilled. 
Indeed, we obtain the effective Hamiltonian 
\begin{equation}
	\begin{aligned}
		H_\text{eff}^{(0)} &=\epsilon\left(\lvert +\rangle\!\langle + \lvert+\lvert -\rangle\!\langle - \lvert\right)+\left(2\epsilon+U\right)\lvert 11\rangle\!\langle 11 \lvert\\	
		&+\frac{2\Gamma}{\pi} \bar{B}(-\epsilon) \left(\lvert 00 \rangle\!\langle 00 \lvert-  \lvert +\rangle\!\langle + \lvert\right)\\
	&+\frac{2\Gamma}{\pi} \bar{B}(-\epsilon-U)\left(\lvert - \rangle\!\langle - \lvert -  \lvert 11\rangle\!\langle 11 \lvert\right).
		\label{eq:upb_Heff}
	\end{aligned}
\end{equation}
Here, $\bar{B}(\epsilon)= \sum_{s=L,R} B_s(\epsilon )/2$ is the average of the function $B_s(\epsilon)$  that arises in the Lamb shift.
This contribution lifts the degeneracy of the dot Hamiltonian caused by \cref{eq:sym_par1}. 
Furthermore, the jump operators are given by
\begin{equation}
	\begin{aligned}
			J_{+ s}^{(0)} &\!=\!\sqrt{2\Gamma}\left[ \sqrt{f_s(\epsilon)}\lvert + \rangle\!\langle 00\lvert +\sqrt{f_s(\epsilon+U)}\lvert 11 \rangle\!\langle -\lvert \right]\!, \\
		J_{- s}^{(0)} &\!=\!\sqrt{ 2\Gamma}\left[\sqrt{1-f_s(\epsilon)}\lvert 00\rangle\!\langle + \lvert+\sqrt{1-f_s(\epsilon+U)}\lvert - \rangle\!\langle 11 \lvert \right]\!,
		\label{eq:upb_jump}
	\end{aligned}
\end{equation}
where $f_s(\epsilon)=\{ 1 + \exp{[(\epsilon - \mu_s)/T_s]} \}^{-1}$ denotes  the Fermi distribution in lead $s$. Further details can be found in \cref{app:Lindblad}.

\subsection{Strong symmetry implications for dynamics} \label{sec:strong}

In the bonding/anti-bonding basis for the dots, for parameters chosen as in \cref{eq:sym_par0,eq:sym_par1},  the  Hamiltonian separates into two distinct sectors which are not coupled by the tunneling processes (see \cite{Li2019,Schaller2009,Kashcheyevs2009}). 
Any initial state of the eigenspace corresponding to the eigenvalue $1$ (or the eigenvalue $-1$) of the swap operator, that is, in the subspace of  $\vert 00\rangle$ and $\lvert + \rangle$ (or the subspace of $\lvert - \rangle$ and $\lvert 11\rangle$), remains supported there at all times.
\\

The dynamics preserves the eigenspaces of $S$ due to the block-diagonal structures of the effective Hamiltonian in \cref{eq:upb_Heff} and the jump operators in \cref{eq:upb_jump}.
Due to the strong symmetry, the parallel dots split into two independent two-dimensional systems in the 
subspaces of  $\vert 00\rangle$ and $\lvert + \rangle$, and of $\lvert - \rangle$ and $\lvert 11\rangle$. 

Indeed, $J_{+ s}$ and $J_{-s}$ facilitate classical transitions between  $\lvert  00 \rangle\!\langle 00 \lvert$ and $\lvert + \rangle\!\langle + \lvert$ at the respective rates 
$2 \Gamma  f_s(\epsilon)$ and $2\Gamma[1- f_s(\epsilon)]$, and between  $\lvert  - \rangle\!\langle -\rvert$ and $\lvert 11 \rangle\!\langle 11\rvert$ at the respective rates $2 \Gamma f_s(\epsilon+U)$ and $2 \Gamma [1-f_s(\epsilon+U)]$, while $H_\text{eff}$ does not contribute.  In fact, these dynamics can be obtained as the Pauli rate dynamics in the bonding-anti-bonding basis. In contrast, a Pauli approach in the local basis (\cref{app:Pauli_stability}) fails \cite{Li2019}, in particular not capturing the degeneracy of stationary states.

The two  stationary states  in the eigenspaces of $S$ are
\begin{equation}
	\begin{aligned}
		\rho^\text{ss}_1  & = [1-\bar{f}(\epsilon)]\lvert 00\rangle\!\langle 00\lvert +\bar{f}(\epsilon)\lvert +\rangle\!\langle + \lvert, \\
		\rho^\text{ss}_2 & =  [1-\bar{f}(\epsilon+U)]\lvert -\rangle\!\langle -\lvert +\bar{f}(\epsilon+U) \lvert 11\rangle\!\langle 11 \lvert,
	\end{aligned}
	\label{eq:unp_stationary}
\end{equation}
where $\bar{f}(\epsilon) = \sum_{s=L,R} f_s (\epsilon )/2 $.
While we have limited ourselves to considering initial states symmetric with respect to $N_\text{PD}$, there are no other stationary states for $U\neq 0$.  
The two stationary states resemble the two degenerate ground states in the corresponding equilibrium system, which exhibits a quantum critical point in the zero-temperature limit \cite{Kashcheyevs2009}.

The stationary currents from the left lead corresponding to the stationary states in \cref{eq:unp_stationary} are
\begin{equation}
	\begin{aligned}
		I_1&= \Gamma \left[f_L(\epsilon)-f_R(\epsilon)\right], \\
		I_2&=   \Gamma \left[f_L(\epsilon+U)-f_R(\epsilon+U)\right].
	\end{aligned}
	\label{eq:unp_current}
\end{equation}
Identical currents are found only if the Coulomb interaction vanishes ($U=0$), or the Fermi distributions for the two leads are the same ($T_L=T_R$, and $V_B=0$). In the former case, the connected configurations feature the same energy difference, such that the system behaves as two identical independent single dots, see also \cref{app:perturbation_0}. In the latter case, the particle currents must asymptotically vanish as there is no directionality induced in the dot system with the leads at equilibrium.
In the limit of infinite Coulomb interaction ($U\rightarrow\infty$),  $f_s(\epsilon+U)\rightarrow 0$, and therefore $I_2\rightarrow 0$. The dynamics between $\lvert - \rangle\!\langle -\lvert$ and $\lvert 11 \rangle\!\langle 11 \lvert$ corresponds then to a decay towards the lower energy state $\lvert-\rangle\!\langle -\lvert$, so it becomes stationary while $\lvert 11\rangle\!\langle11\lvert$ is prohibited energetically  [\cref{eq:unp_stationary}]. But the asymptotic current $I_1$ remains unchanged as it is independent from $U$.

The two stationary states in \cref{eq:unp_stationary} fix the choice of the right eigenmatrices for  two zero eigenvalues of the unperturbed Liouvillian. The corresponding left eigenmatrices are given by the projections on their supports,
\begin{equation}
	\begin{aligned}
		&P_{1}=\lvert 00\rangle\!\langle 00\lvert +\lvert +\rangle\!\langle + \lvert=\frac{\mathds{1}+S}{2},\\
		&P_{2}=\lvert -\rangle\!\langle -\lvert +\lvert 11\rangle\!\langle 11 \lvert=\frac{\mathds{1}-S}{2},
	\end{aligned}
	\label{eq:unp_proj}
\end{equation}
These determine  the asymptotic state for a general initial state $\rho_\text{PD}(0)$ as  $p_1 	\rho^\text{ss}_1 + p_2	\rho^\text{ss}_2 $, where $p_j =\text{Tr}[P_j \rho_\text{PD}(0)]$, $j=1,2$, and thus the stationary current to $p_1 I_1+p_2 I_2$. 

Next to the stationary states, there are two decay modes corresponding to the classical dynamics with the degenerate eigenvalues 
\begin{align}
\lambda_5^{(0)}=\lambda_6^{(0)}=-4\Gamma.
\label{eq:lambda56}
\end{align}
The other two modes describe the decay of coherences in the bonding/anti-bonding basis with the eigenvalues
\begin{align} 
\lambda_3^{(0)}=\left[\lambda_4^{(0)}\right]^*& =-2\Gamma  [1-\bar{f}(\epsilon)+\bar{f}(\epsilon+U)]	\notag\\
&+i \frac{2\Gamma}{\pi }[\bar{B}(-\epsilon) + \bar{B}(-\epsilon - U)],
\label{eq:lambda34}
\end{align}
where the oscillation frequency arises  from the Lamb shift. For the eigenmatrices, see \cref{app:perturbation_0}. \\

\subsection{Breaking of strong symmetry and metastability} 
\label{sec:perturbed}

We now consider perturbations in the dynamical parameters that break the strong swap symmetry.  As a consequence,  the two-fold degeneracy of the zero-eigenvalue of the Liouville operator is lifted, and a unique stationary  state arises together with a new timescale in the dynamics for the system's final relaxation. Using non-Hermitian perturbation theory \cite{Kato}, we investigate those aspects of the dynamics with a focus on how the  current is affected.\\

We examine perturbations that break the degeneracy of the dot energy levels as
\begin{align}
	\epsilon_1 &= \epsilon - \delta\epsilon,  \quad \epsilon_2 = \epsilon + \delta\epsilon,
	\label{eq:per_epsilon}
\end{align}
 and for definiteness choose to alter the tunneling rates as
\begin{equation}
	\begin{aligned}
	\Gamma_{1L} &=\Gamma -  \delta \Gamma,  \quad  &&\Gamma_{1R} = \Gamma+  \delta \Gamma,\\
	\Gamma_{2L}  &= \Gamma+  \delta \Gamma,  \quad  &&\Gamma_{2R}= \Gamma -  \delta \Gamma.
	\label{eq:per_Gamma}
\end{aligned}
\end{equation}
In this work, we focus on small perturbations in the sense that $\delta \epsilon, \delta \Gamma   \ll \Gamma$, which also implies that $\delta \epsilon \ll T$.
This allows us to exploit non-Hermitian perturbation theory to characterize the eigenvalues and eigenmatrices of the perturbed Liouvilian,
\begin{align}\label{eq:per_master}
	\mathcal{L}  =   \mathcal{L}^{(0)} + \mathcal{L}^{(1)} + \mathcal{L}^{(2)} + \dots.
\end{align}
Above, we have expanded $\mathcal{L}$ in the perturbation parameters, where the superscript indicates the order of the perturbation. 
We consider corrections to $\mathcal{L}$ within the stationary state manifold of $\mathcal{L}^{(0)}$, which consists of probabilistic mixtures of the stationary states in \cref{eq:unp_stationary}. The focus of the following discussion is on the  physical aspects; technical details can be found in \cref{app:perturbation}, see also Supplemental Materials in Refs.\,\cite{Macieszczak2016,Macieszczak2021}.

The first-order correction is necessarily zero,   due to the effective classicality of the manifold of the stationary states of $\mathcal{L}^{(0)}$. The second-order correction corresponds to the classical  dynamics of the probabilistic mixtures of the unperturbed system's stationary states in \cref{eq:unp_stationary},
\begin{equation}
	\frac{d}{dt} 
	\left[\begin{array}{cc}
	p_1(t) \\
	p_2(t)
\end{array}\right]
	=
	\begin{pmatrix}
	-\gamma_1 & 	\phantom{-}\gamma_2\\
		\phantom{-}\gamma_1 & -\gamma_2
	\end{pmatrix}
	\left[\begin{array}{cc}
		p_1(t) \\
		p_2(t)
	\end{array}\right],
	\label{eq:eff_dynamics}
\end{equation} 
where $p_{1,2} (0)= \mathrm{Tr}[P_{1,2}\rho_\text{PD}(0)]$, so that  $p_1(t)+p_2(t)=1$ as $P_1+P_2=\mathds{1}$ and the dynamics in \cref{eq:eff_dynamics} conserves the total probability. The decay rates $\gamma_{1,2}$  can be found by expressing the second-order corrections to the reduced dynamics in the basis of Eqs.~\eqref{eq:unp_stationary} and~\eqref{eq:unp_proj}, and they are quadratic functions of $\delta \epsilon$ and $\delta \Gamma$.
We now exploit this result in two ways. \\

First, the stationary probability distribution for the dynamics in \cref{eq:eff_dynamics}, $p^\text{ss}_1 = \gamma_2/(\gamma_1 + \gamma_2)$ and  $p^\text{ss}_2 = \gamma_1/(\gamma_1 + \gamma_2)$,
gives for the stationary state
\begin{align}
 \rho_\text{PD}^\text{ss} =p^\text{ss}_1 \rho^\text{ss}_1 +p^\text{ss}_2  \rho^\text{ss}_2 + \dots.
 \label{eq:per_ss}
\end{align} 
The higher order corrections, indicated by $...$, are of the first order. They can be understood to arise as the corrections to the structure of the two states corresponding to the stationary states in the unperturbed limit, which now constitute the two metastable phases. 

It is important to note that because $\gamma_1$ and $\gamma_2$ depend on two rather than a single perturbation, the stationary state depends on the perturbations already in lowest order, but only via their ratios, which can be seen as a competition between the metastable phases.
This leads to distinct stationary current values when perturbations are varied,  which will be discussed in more detail later, see also \cref{sec:sensing}. Indeed, the stationary current is 
\begin{align}
    I = p_1^\text{ss} I_1 + p_2^\text{ss} I_2+...,
    \label{eq:per_current}
\end{align}
where the corrections are at least of the second-order.

Finally, since $\gamma_{1}\neq \gamma_2$ in general, the stationary state will feature coherences in the local basis. Therefore, the approximation of the dynamics by a Pauli rate equation in that basis will fail, see also \cref{app:Pauli_stability}.\\

Second, these results can be used to understand the long-time dynamics. For times longer than the initial relaxation, $ t \gg -1/\text{Re}(\lambda_{3,4})$, which in the leading order is determined by the eigenvalues $\lambda_{3,4}^{(0)}$ of the unperturbed dynamics, the dynamics can be understood  as follows.

The two-fold degeneracy of the zero-eigenvalue is now lifted with the second eigenvalue
\begin{align}\label{eq:per_lambda2}
\lambda_2=-(\gamma_1+\gamma_2)+...,
\end{align} 
so that a new timescale emerges in the dynamics. This low-lying eigenvalue is accompanied by the right and left eigenmatrices 
\begin{equation}\label{eq:per_decay}
\begin{aligned}
    R_2 &=     \rho^\text{ss}_1-\rho^\text{ss}_2+...,\\
    L_2 &= p^\text{ss}_2 P_1- p^\text{ss}_1 P_2+....,
\end{aligned}
\end{equation}
where $P_{1}$ and $P_2$ are defined in \cref{eq:unp_proj} and corrections are at least of  the first order. 
For times $t$ such that $t|\delta \lambda_2| \ll 1$, with $\delta \lambda_2$ denoting the corrections in \cref{eq:per_lambda2}, which are at least fourth order, the state of the dots (see \cref{eq:evolution_meta}) can be approximated as a probabilistic mixture of  the states corresponding to stationary states in the unperturbed system,
\begin{align}
	\rho_\text{PD}(t) = p_1(t) \rho^\text{ss}_1 + p_2(t)\rho^\text{ss}_2 + \dots,
	\label{eq:prop_mixture}
\end{align} 
with the corrections of the first order. 
Access to later times can be gained by including higher order corrections in \cref{eq:eff_dynamics}
(or modifying the generator even further \cite{Burgarth2021}). The numerical methods introduced in Ref.\, \cite{Macieszczak2016,Rose2016} allow for the study of the long-time dynamics to all orders, simply by considering the low-lying part of the spectrum of the Liouvillian $\mathcal{L}$, as obtained by its diagonalization; for a short summary, see  \cref{app:num_metastability}.

\begin{figure}[t!]
	\centering
	\includegraphics[width =0.95 \linewidth]{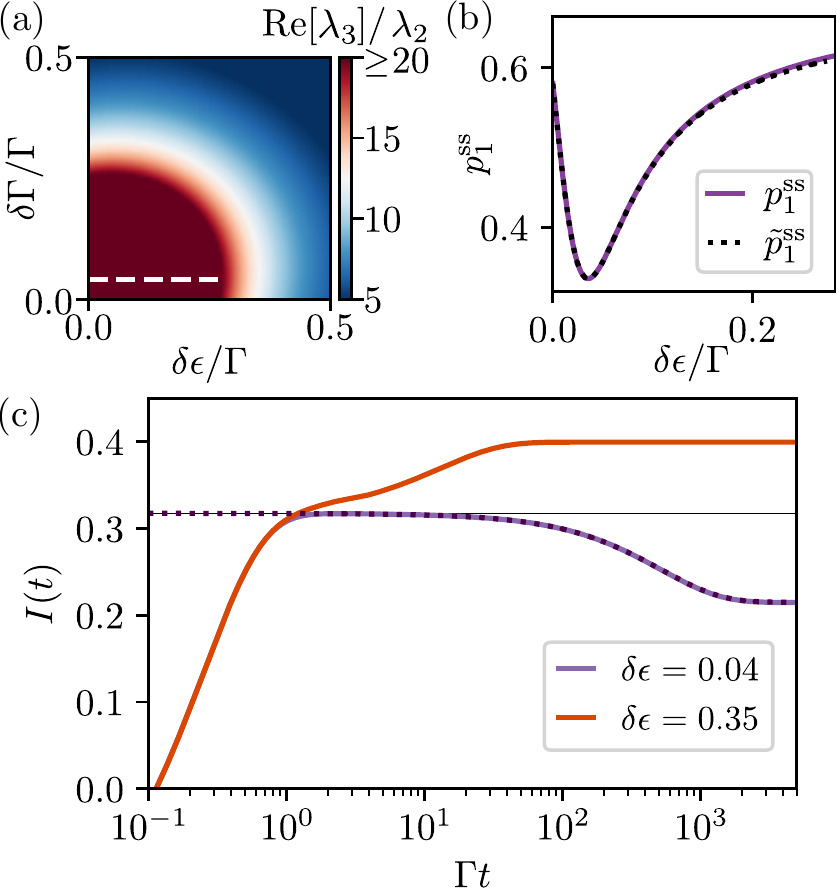}
	\caption{
		(a) $\text{Re}(\lambda_3)/\lambda_2$  as a function of $\delta \epsilon, \delta \Gamma$. We (somewhat arbitrarily) define the parameter regime where metastability occurs by $\text{Re}(\lambda_3)/\lambda_2 > 20$. (b) Stationary probabilities of \cref{eq:eff_dynamics} (purple line) well approximate  the numerical decomposition into metastable phases $\tilde{p}_1$ (black dotted line) for constant $\delta \Gamma/\Gamma = 0.04$ [shown as the white dashed line in (a)], see \cref{app:num_metastability}). (c) The transient current  for $\delta \epsilon /\Gamma= 0.04$ (purple line) and $\delta \epsilon/\Gamma = 0.35$ (orange line). For a small perturbation $\delta \epsilon$, the transient dynamics shows a plateau of approximately constant current corresponding to \cref{eq:current_metastable} (gray line), which at later times follows the evolution of \cref{eq:eff_dynamics} (purple dashed line).  In both cases, $\delta \Gamma/\Gamma = 0.04$, and the initial state is the fully mixed state  $\rho_\text{PD}(0) = \mathds{1}/4$ in the local basis.  Other parameters are chosen as in \cref{fig:model}(b).}
	\label{fig:metastable_current}
\end{figure}

For times $t$ such that $-t\lambda_2 \ll 1$, the effective dynamics in \cref{eq:eff_dynamics} can be neglected and the system is approximately stationary, i.e., metastable,
\begin{align}
    \rho_\text{PD}(t) = p_1(0)\rho^\text{ss}_1+p_2(0)\rho^\text{ss}_2+...,
    \label{eq:metastable_state}
\end{align}
with the leading contribution given by the asymptotic state of the unperturbed dynamics. In this metastable regime, the unperturbed system's stationary states take the role of metastable phases, and the system can be in any probabilistic mixture of these depending on the initial state \cite{Macieszczak2016,Rose2016}. As a consequence, a whole range of approximately constant average currents  can be supported in this regime, 
\begin{align}
	I(t) = p_1(0)I_1+p_2(0)I_2+....
	\label{eq:current_metastable}
\end{align}
In contrast to \cref{eq:per_current},  the current values are in the leading order determined by the initial system state and thus independent from the perturbations.  \\

We  demonstrate our analytical findings of the parallel dots' dynamics for concrete parameter choices in \cref{fig:metastable_current}.
In \cref{fig:metastable_current}(a), we characterize perturbation strengths for which the perturbative approach is applicable, with the metastability criterion  \cite{Macieszczak2016,Rose2016} $\text{Re}(\lambda_3)/\lambda_2 \gg 1$. In this regime, the stationary state is well approximated by the zeroth order terms in \cref{eq:per_ss}. To show this, the stationary probabilities of \cref{eq:eff_dynamics} are plotted as a function of the detuning $\delta \epsilon$ in \cref{fig:metastable_current}(b). They are compared with the decomposition into two metastable phases constructed numerically to all orders, see also ~\cref{app:num_metastability}.  Changing the dot energy perturbation while keeping a fixed tunneling rate perturbation,  the distribution varies non-monotonically. In turn the stationary current is impacted, and  can be suppressed when the metastable phase with the vanishing current (due to a large Coulomb interaction) predominantly contributes.

The transient dynamics in the regime where the system is metastable is qualitatively different to where the system does not exhibit metastability. In \cref{fig:metastable_current}(c) the transient current calculated from the full dynamics [\cref{eq:evolution3,eq:current}] is plotted for different choices of the detuning $\delta \epsilon$.
For small perturbations, metastability occurs and
the transient current remains approximately constant over a long time after the initial dynamics,  see \cref{eq:current_metastable}, before it finally evolves into its true stationary value,  see also \cref{eq:per_current}.  That long-time evolution is well approximated by the classical effective dynamics of \cref{eq:eff_dynamics}.
For larger perturbations,
the current evolves towards its stationary value continuously as the perturbative approach breaks down and to capture the dynamics correctly, the full expression for the evolution of \cref{eq:evolution3} is needed.

\begin{figure*}[t!]
	\centering
	\includegraphics{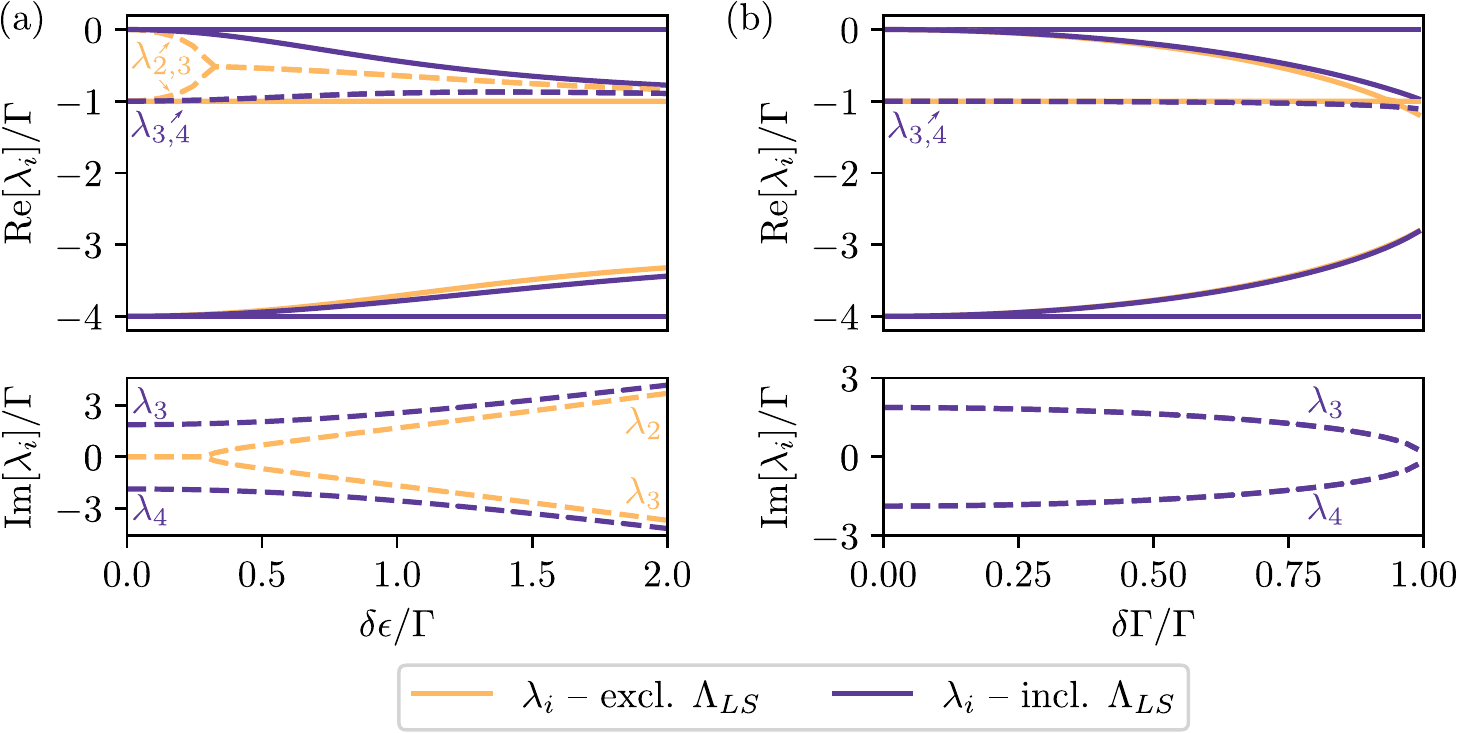}
	\caption{ 
		Real (top panels) and imaginary part (bottom panels) of the spectrum of $\mathcal{L}$ as function of the perturbations $\delta \Gamma$, $\delta \epsilon$ with  the Lamb shift $H_\text{LS}$ included (purple) and excluded (orange).  The solid lines correspond to purely real eigenvalues while the dashed lines represent eigenvalue branches with a non-vanishing imaginary part. (a) Spectrum for varying $\delta \epsilon/\Gamma$ and constant $\delta \Gamma/\Gamma = 10^{-8}$.
		(b) Spectrum for varying $\delta \Gamma/\Gamma$ and constant $\delta \epsilon/\Gamma = 10^{-6}$; 
     all other parameters are chosen as in \cref{fig:model}.}
	\label{fig:spectrum}
\end{figure*}

\subsection{Dynamics beyond small perturbations} \label{sec:spectrum}

Further insight into the system dynamics is provided by the eigenvalue spectrum of the Liouville operator in \cref{fig:spectrum}.

For increasing perturbation strengths, the difference between $\lambda_2$ and $\text{Re}(\lambda_3)$ decreases, so there is no clear separation between the corresponding  decay rates and the metastability is absent. Due to the Lamb shift, the eigenvalues $\lambda_{3,4}$ acquire an imaginary part which scales with the difference of the (renormalized) energy levels of the system, while $\lambda_2$ remains real. Disregarding the Lamb shift leads to a fundamental difference for $\delta \Gamma=0$ and varying  $\delta \epsilon$ [\cref{fig:spectrum}(a)]. In this case,  the spectrum remains real below a threshold in $\delta \epsilon$, above which the eigenvalues $\lambda_2$ and $\lambda_3$ merge and their corresponding eigenvectors are identical, so that the spectrum exhibits an exceptional point \cite{Heiss2004}. For larger $\delta \epsilon$, the merged branches acquire an imaginary contribution forming a complex conjugate pair. The differences between the evolutions with and without the Lamb shift are less pronounced along constant $\delta \epsilon$ but varying $\delta \Gamma$.

For large detuning $\delta\epsilon /\Gamma \gg 1$, we observe the loss of coherence in the system as the real parts of the eigenvalues approach the values $0, -\Gamma, -3\Gamma, -4\Gamma$. These correspond to the eigenvalues of the Pauli rate equation that describes the dynamics of a density matrix assumed diagonal in the local basis \cite{Kirsanskas2017}; see also \cref{app:Pauli_stability}. 

%%%%%%%%%%%%%%%%%%%%%%%%%%%%%%%%%%%%%%%%%%%%%%%%%%%%%%%%%%%%%%%%%%%%%%%

%%%%%%%%%%%%%%%%%%%%%%%%%%%%%%%%%%%%%%%%%%%%%%%%%%%%%%%%%%%%%%%%%%%%%%%
%             SEC: Sensing
%%%%%%%%%%%%%%%%%%%%%%%%%%%%%%%%%%%%%%%%%%%%%%%%%%%%%%%%%%%%%%%%%%%%%%%

\section{Quantum coherence assisted sensing} \label{sec:sensing}

The coherences in the parallel dots lead to a stationary current which changes significantly as the perturbations   $\delta \epsilon$ and $\delta \Gamma$ vary. From Sec.~\ref{sec:perturbed}, in the perturbative regime we understand this as the result of the competition of  two metastable phases, but even for larger perturbations the current remains sensitive to changes in the perturbations \cite{Li2019}. 
We now investigate the possibility of using the parallel dots as a sensor when a parameter quench is detected through its effect on the particle current. 
In particular, we consider sensing a change in the nearby charge distribution which is assumed to lead to a shift in the perturbation of the dot energies $\delta \epsilon$. The change in charge distribution could, for example, be due to a single electron being added to or removed from another nearby quantum dot \cite{Hu2007,Podd2010,Salfi2010}.

We consider a continuous measurement of the current in \cref{eq:current} during a time $\tau$ in order to detect the parameter quench that has occurred at  $t=0$. The statistics of such a measurement can be accessed using full counting statistics \cite{Emary2009,Schaller,Flindt2008}, see  Appendix~\ref{app:counting_statistics}.  

\begin{figure*}
	\centering
	\includegraphics[width = \linewidth]{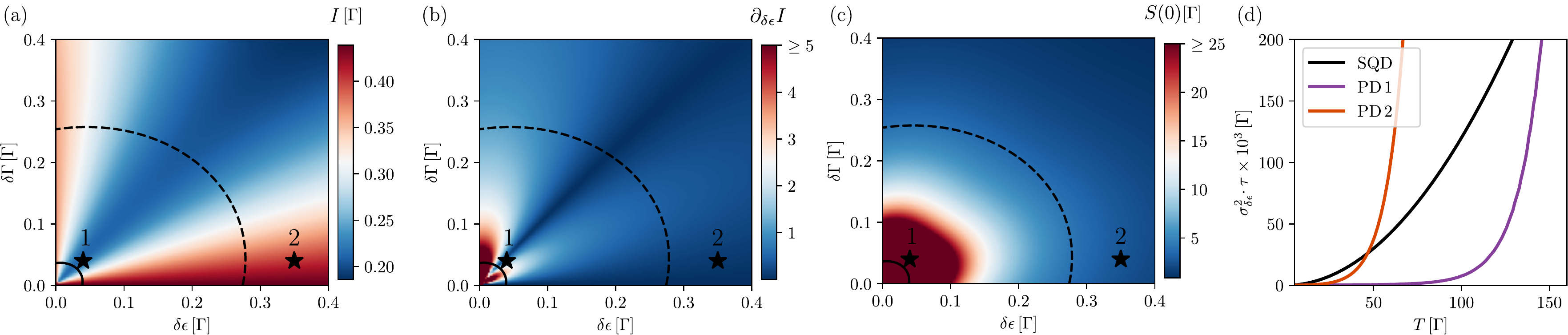}
	\caption{(a) The stationary  current,  (b) absolute value of the signal rate $\lvert \partial_{\delta \epsilon}I\rvert $ and (c) fluctuation rate as functions of $\delta \Gamma$ and $\delta \epsilon$. All other parameters are chosen as indicated by the marker in \cref{fig:model}(b). The black dashed line in (a)--(c) marks the approximate border where metastability occurs and the solid black contour marks where $-\lambda_2/\Gamma > 10^3$. (d) Temperature dependence of $\sigma_{\delta \epsilon}^2$ of the parallel dots (PD) for $\delta\epsilon = \delta \Gamma = 0.04 \Gamma$ (purple, marker 1),   $\delta\epsilon =  0.35, \delta \Gamma = 0.04 \Gamma$ (orange, marker 2). These points are indicated by stars in (a)--(c). The error $\sigma_{\epsilon}^2$ for a single quantum dot (SD) in the ideal configuration is plotted in black. }
	\label{fig:sensor}
\end{figure*}

The experimentally feasible measurement time $\tau$ sets a lower bound on the perturbation strength we consider.
Indeed, the typical measurement time for charge detection in quantum dots is $\tau \sim \mu\text{s}$ \cite{Reilly2007,Cassidy2007}, while a typical value for tunneling rate sets $\Gamma^{-1} \sim$ ns. 
We assume that not only the initial but also the final relaxation takes place within the measurement time, and we therefore calculate all quantities in \cref{fig:sensor} in the stationary state. 
This means that the system cannot be too far into the regime where relaxation becomes extremely slow, which puts some lower bound on the perturbation strength, see \cref{fig:sensor}.

In \cref{fig:sensor}(a), the stationary current as a function of the two perturbations $\delta \epsilon, \delta \Gamma$ clearly depends only on the perturbation ratio, but not visibly on the perturbation strength.  For the regime where the system exhibits metastability this is expected  in terms of the dependence of the current on $p_{1}^\text{ss}$, see \cref{eq:per_current}. 

The signal of the (time-)integrated current is asymptotically linear in time with the rate equal to the response of the stationary current to a change in $\delta \epsilon$,  which is shown in \cref{fig:sensor}(b).  In contrast to the current it depends on the perturbation strength and diverges as that is reduced.  
In the perturbative regime, this response is dominated by the change in the stationary probabilities,
\begin{align}\nonumber
	\partial_{\delta \epsilon} I&= \partial_{\delta \epsilon}p_1^\text{ss} (I_1-I_2)+...\\&=\frac{ \gamma_2\partial_{\delta \epsilon}\gamma_1 -\gamma_1 \partial_{\delta \epsilon} \gamma_2}{(\gamma_1+\gamma_2)^2}\,(I_1-I_2)+...,
	\label{eq:signal}
\end{align}
where we used $ \partial_{\delta \epsilon} p_1^\text{ss}=- \partial_{\delta \epsilon} p_2^\text{ss}$ as  $p_1^\text{ss}+p_2^\text{ss}=1$. Indeed, $\partial_{\delta \epsilon} \gamma_j$ is of the first order while $\gamma_j$ is of the second order for $j=1,2$, so that $\partial_{\delta \epsilon} I$ diverges with the inverse of the perturbations.

This is true except for when $\partial_{\delta \epsilon}  p_1^\text{ss}=0$, which occurs in two cases. First, for the perturbation at $\delta \Gamma = 0$ since $p_1^\text{ss}$ is then independent from $\delta \epsilon$. Second, when the perturbation ratio corresponds to the minimal current in  \cref{fig:sensor}(a). In those cases, the signal rate appears to actually vanish [as  the corrections to the stationary current in \cref{eq:per_current} are of the second-order,  the signal rate is then of the first-order].

The variance of the integrated current is asymptotically linear in time with the rate $S(0)$ equal to the zero-frequency noise. 
The divergence of the fluctuation rate is evident in \cref{fig:sensor}(c) and agrees with the behaviour of the Fano factor $F = S(0)/I$ observed for  $\delta \epsilon, \delta \Gamma \rightarrow 0$ in Ref. \cite{Schaller2009}. 
Indeed, for smaller perturbations, metastability arises leading to long-lived correlations in the current so that its fluctuation rate becomes large \cite{Macieszczak2016b,Schaller2009}
\begin{align}\nonumber
	S(0)&= \frac{2}{\lambda_2} p_1^\text{ss} (1-p_1^\text{ss}) (I_1-I_2)^2+...\\&= \frac{2 \gamma_{1}\gamma_2}{(\gamma_{1}+\gamma_{2})^3} (I_1-I_2)^2+...
	\label{eq:noise} 
\end{align}
For fixed $\gamma_1+\gamma_2$, the smallest multiplicative factor corresponds to the stationary probability $p_1^\text{ss}$ being minimal or maximal.
Outside the parameter regime where metatability occurs, $S(0)$ saturates to a constant value.

Estimation errors are determined via the standard error propagation formula as the measurement variance rescaled by the square of its signal. As the measurement time $\tau$ is assumed much longer than the relaxation time, the variance is dominated by $\tau S(0)$,  and the signal by $\tau \partial_{\delta\epsilon} I$.  Therefore, the error is given by
\begin{align}
	\sigma_{\delta\epsilon}^2 &=\frac{1}{\tau} \frac{S(0)}{ \left(\partial _{\delta\epsilon} I \right)^2}\label{eq:error_sensor} \\
	&\approx \frac{2}{\tau} \frac{ \gamma_{1}\gamma_2\left(\gamma_{1}+\gamma_{2}\right)}{\left(\gamma_2\partial_{\delta\epsilon} \gamma_1 -\gamma_1 \partial_{\delta\epsilon} \gamma_2\right)^2}\notag,
\end{align}
where the second line holds for the perturbative regime where metastability occurs. 
We see that the divergence of the signal rate in \cref{eq:signal} and the fluctuations rate in \cref{eq:noise} cancel out, and the difference in currents  between the metastable phases simplifies as well. 

For the best sensing setup, the errors in \cref{eq:error_sensor} should be minimised by the choice of the perturbation values before the quench. This corresponds to a trade-off between  the fast divergence of the signal  and the slow divergence of the fluctuations, which is non-trivial as, for fixed $\gamma_1+\gamma_2$, the slowest divergence of the fluctuations occurs exactly when the current is maximal or minimal and the divergence of the signal is actually absent. 

We benchmark the performance of the parallel dots sensor with a single quantum dot setup for charge sensing \cite{Biercuk2006,Podd2010,Bauerle2018}, where a change in the charge distribution is assumed to affect the dot energy $\epsilon$.  
The parameters for the single quantum dot are chosen such that it operates at a conductance peak where sensitivity is maximal, see  \cref{app:Pauli_stability}. The width of the conductance peak depends on the temperature where $\partial_\epsilon I \propto 1/T$, while $S(0)$ is independent of $T$.
Consequently the errors $\sigma_{ \epsilon}^2$ show a quadratic temperature dependence.  
The advantage of the parallel dots system operating as a sensor is that it is not limited by temperature in contrast to the single dot setup. In \cref{fig:sensor}(d) the temperature dependence of the error in \cref{eq:error_sensor} is shown for both setups. For the double quantum dot, the parameters are chosen close to the minimal value of the current (marker 1 in \cref{fig:sensor}), where the signal rate is numerically observed to diverge fast, and small $\delta \epsilon$ which results in the corresponding errors $\sigma_{\delta \epsilon}^2$ remaining small over a wide range of temperatures.  For example,  at $T/\Gamma \sim 60$ the ratio of error for single and parallel dots $\sigma^2_{\epsilon}/\sigma^2_{\delta \epsilon}\sim 55$. 
In the limit $T/\Gamma\to0$, the error of parallels dots approaches a constant value. 

The parameter region in which the parallel dots can be operated as a sensor is not limited to the chosen set of parameters indicated in the stability diagram in \cref{fig:model}(b). In fact, the trade-off between the noise and the signal anywhere between high conductance lines for positive bias voltages shows similar behaviour,  see  \cref{app:stab_dia}.

%%%%%%%%%%%%%%%%%%%%%%%%%%%%%%%%%%%%%%%%%%%%%%%%%%%%%%%%%%%%%%%%%%%%%%%

%%%%%%%%%%%%%%%%%%%%%%%%%%%%%%%%%%%%%%%%%%%%%%%%%%%%%%%%%%%%%%%%%%%%%%%
%                  SEC: Conclusion
%%%%%%%%%%%%%%%%%%%%%%%%%%%%%%%%%%%%%%%%%%%%%%%%%%%%%%%%%%%%%%%%%%%%%%%
%

\section{Conclusions}

We have analyzed the transient dynamics and non-equilibrium transport properties of two interacting parallel quantum dots coupled to macroscopic leads. 

A swap symmetry between the dots is present when the dots energies are degenerate and tunneling rates to both dots are identical.  This parity-like symmetry leads to the existence of two stationary states distinguishable by their currents values.  In equilibrium, this parity-like symmetry translates to a $SU(2)$ symmetry of a pseudo-spin \cite{Lee2007}, and its diverging susceptibility indicates a quantum critical point \cite{Kashcheyevs2009}. 
Perturbations of dot energies and tunneling rates introduce metastability into the dots dynamics, and long-time dynamics  towards a unique stationary state arises with a rate that is quadratic in the perturbations. Crucially, the stationary state depends already in the leading order on the ratio of the perturbations in dot energies and tunneling rates. This leads to the diverging signal (change in the stationary current in response to a small change in the perturbations). Since the current fluctuations also diverge in this limit, we found that the signal to noise ratio remains finite, but dependent on the perturbation ratios. In the context of charge sensing, a comparison with a single dot showed that the parallel dots may perform significantly better. 

While the dynamics of the parallel dots was considered in this work as GKLS dynamics beyond the secular approximation, the discussed aspects were directly connected to those of the Hamiltonian dynamics of the dots and the leads, especially in the context of  symmetry breaking. Therefore, our results should be qualitatively valid for other approximations for the dot dynamics \cite{Li2019}.

The physics observed in this work crucially depends on coherent dynamics. It would therefore be interesting to investigate how the results change when including additional decoherence mechanisms, for example due to charge fluctuations in the environment.
 Additionally, how strong-correlation signatures in the conductance \cite{Meden2006,Kashcheyevs2007} are altered in the non-equilibrium setup, and how the corresponding current noise influences the sensitivity in a potential sensing application, remain open questions for future studies.

%%%%%%%%%%%%%%%%%%%%%%%%%%%%%%%%%%%%%%%%%%%%%%%%%%%%%%%%%%%%%%%%%%%%%%%
%               Acknowledgements
%%%%%%%%%%%%%%%%%%%%%%%%%%%%%%%%%%%%%%%%%%%%%%%%%%%%%%%%%%%%%%%%%%%%%%%
%

\begin{acknowledgments}
We would like to thank R.\ Seoane Souto and V.\ Kashcheyevs for fruitful discussions. 
We acknowledge financial support from the Swedish Research Council (VR), the Knut and Alice Wallenberg Foundation (project 2016.0089),  the Wallenberg Center for Quantum Technologies (WACQT), and from NanoLund.
K.M. acknowledges the support from the Henslow Research Fellowship and, as a visiting researcher, from the Department of Physics, University of Cambridge.
\end{acknowledgments}

%%% Bibliography %%%%%%%%%%%%%%%%%%%%%%%%%%%%%%%%%%%%%%%%%%

\bibliography{references}

%%%%%%%%%%%%%%%%%%%%%%%%%%%%%%%%%%%%%%%%%%%%%%%%%%%%%%%%%%%%%%%%%%%%%%%
%                   APPENDICES
%%%%%%%%%%%%%%%%%%%%%%%%%%%%%%%%%%%%%%%%%%%%%%%%%%%%%%%%%%%%%%%%%%%%%%%
%

\section*{Appendixes}
\appendix
\section{Master equation} \label{app:Lindblad}

We derive the equation of motion for the reduced density matrix $\rho_\text{PD}$ of the quantum dots 
of GKLS form as stated in the main text in \cref{eq:evolution}.

\subsection{Effective Hamiltonian} \label{app:Lindblad_Heff}

The dot Hamiltonian of \cref{eq:HQD}, in its eigenbasis ordered as $|00\rangle$, $|10\rangle$, $|01\rangle$ and $|11\rangle$, takes the following form,
\begin{align}
	\label{app_eq:HQD}
		H_\text{PD}& = 
		\begin{pmatrix}
			 0  &0 & 0 & 0 \\
			0 & \epsilon_1 & 0 & 0  \\
			0 & 0 &  \epsilon_2& 0 \\
			0 & 0 & 0 & \epsilon_1 + \epsilon_2 + U
		\end{pmatrix}.
\end{align}

To find the effective Hamiltonian, we need to determine the Lamb shift    $H_\text{LS}$  describing the renormalization of the system's eigenenergies due to the coupling to the leads. 
The latter is beyond the secular approximation given by \cite{Ptaszy2019}
\begin{align}
	  H_\text{LS} = \frac{1}{2} \sum_{lmn} \sum_{a b} & \left[S_{a b}(\omega_{ml}) + S_{ab}(\omega_{mn})\right]\notag\\
	& \times
	X^{(a)}_{lm} X^{(b)}_{mn} \lvert l\rangle\!\langle n \lvert.
\end{align}
The operators $X^{(a)}$, $X^{(b)}$, $a,b = 1,\dots,4$, correspond to the physical processes generated in the dots by the tunneling Hamiltonian in \cref{eq:HT}, i.e., 
\begin{equation}
	\begin{aligned}
     &X^{(1)}  = d^\dagger_1, && X^{(2)} = d^\dagger_2, \\
     &X^{(3)}  = d_1, && X^{(4)} = d_2.
     \label{eq:operator_sys}
\end{aligned}
\end{equation}
Furthermore, $l,m,n$ label the eigenbasis  of $H_\text{PD}$, while  $\omega_{mn} = E_m - E_n$ denote the corresponding energy differences.
Finally,  $S_{{\alpha}{\beta}}(\omega)$ is determined by the odd Fourier transform, 
\begin{align}
	   i S_{a b}(\epsilon) &= \int_{-D}^D d\tau \sgn{(\tau)} C_{ab}(\tau) e^{i\omega \tau}\notag\\
	  & = \frac{i}{2\pi} \mathcal{P}\int_{-D}^{D} d\omega\, \frac{ C_{ab}(\omega)}{\epsilon - \omega},
\end{align}
where $D$ is the bandwidth. 
The function   $C_{ab}(\omega) =\int d\tau\, C_{ab}(\tau) e^{i\omega \tau}$, is in turn the Fourier transform of the lead correlation function
\begin{align}
	    C_{ab}(t-t') = \text{Tr}\!\left[Y^{(a)}(t) Y^{(b)}(t') \rho_L\right].
\end{align}
Here, $\rho_L$ is the state of electrons in the leads, assumed to be given by the grand canonical ensemble and the evolution in the interaction picture [i.e., with the lead $H_L$ in~\cref{eq:HL}]. The operators $Y^{(i)}$ are defined as  [cf.~\cref{eq:HT}]
\begin{equation}
	\begin{aligned}
     &Y^{(1)}  = \sum_{s=L,R}\sum_k t^*_{1,ks} c_{sk}, &&Y^{(2)}  = \sum_{s=L,R}\sum_k t^*_{2,ks} c_{sk},\\
     & Y^{(3)} = \sum_{s=L,R}\sum_k t_{1,ks} c^\dagger_{sk}, &&Y^{(4)} = \sum_{s=L,R}\sum_k t_{2,ks} c^\dagger_{sk}.
     \label{eq:operator_bath}
\end{aligned}
\end{equation}
Equations \eqref{eq:operator_sys} and \eqref{eq:operator_bath} are a standard choice when treating non-equilibrium transport setups \cite{Schaller}. Note that due to the conservation of the electron number in the leads, i.e., $[N_s,H_L]=0$ [cf.~ \cref{eq:HL}], and the initial state such that $[N_s,\rho_L]=0$; we have $C_{12}(\tau)=C_{21}(\tau)=C_{34}(\tau)=C_{43}(\tau)=0$. 

In the continuum limit, with the assumption of tunneling amplitudes being independent from momentum $k$,  the Fourier transforms of non-zero correlation functions  are given by
\begin{align}
	C_{13}(\omega) &= 2\pi\nu \sum_{s=L,R}\lvert t_{1s}\rvert^2 \left[1 - f_s\left(\omega \right)\right],\notag\\
	C_{14}(\omega) & = 2\pi \nu \sum_{s=L,R} t^*_{1s} t_{2s} \left[1 - f_s\left(\omega \right)\right]= C_{23}^*(\omega) ,\notag\\
	C_{24}(\omega) &= 2\pi\nu \sum_{s=L,R}\lvert t_{2s}\rvert^2 \left[1 - f_s\left(\omega \right)\right],\\
		C_{31}(\omega) &= 2\pi\nu \sum_{s=L,R}\lvert t_{1s}\rvert^2  f_s\left( - \omega \right),\notag\\
		C_{32}(\omega) &= 2\pi\nu \sum_{s=L,R} t_{1s} t^*_{2s}  f_s\left(- \omega \right)= C_{41}^*(\omega) ,\notag\\
		C_{42}(\omega) &=2\pi \nu \sum_{s=L,R}\lvert t_{2s}\rvert^2  f_s\left( - \omega \right)\notag.
\end{align}

Here, $\nu$ is the density of states (assumed constant) and the Fermi distribution
\begin{align}\label{eq:Fermi}
    f_s (\omega) = \frac{1}{e^{\left(\omega - \mu_s\right)/T_s} + 1},
\end{align}
with the  temperature $T_s$ and the chemical potential $\mu_s$ for the lead $s$.

 In the limit of a large bandwidth $D\to\infty$, for $S_{{\alpha}{\beta}}(\omega)$ we make use of the principal value integral \cite{Ptaszy2019,Abramowitz,Kirsanskas2017}
  \begin{widetext}
\begin{align}
  &\lim_{D\to \infty}\mathcal{P}\int_{-D}^D d\omega \frac{ f_s(\omega)}{\epsilon - \omega} \approx-\text{Re}\left\{\Psi \left[\frac{1}{2} + i \frac{\beta_s\left(\epsilon - \mu_s\right)}{2\pi}\right] \right\}+ \ln{\left(\frac{D\beta_s}{2\pi}\right)} \equiv B_s(\epsilon),
  \label{eq:approx}
\end{align}
with the inverse temperature $\beta_s$ and the digamma function $\Psi$, to arrive at the Lamb shift Hamiltonian $H_\text{LS}$ with the following non-zero entries

\begin{equation}
\begin{aligned}
	\langle 00 \rvert H_\text{LS} \lvert 00\rangle & = \sum_{j=1,2} 	\sum_{s=L,R}  |t_{js}|^2 B_s (-\epsilon_j), \\
	\langle 10 \rvert H_\text{LS} \lvert 10\rangle & =  -	\sum_{s=L,R}  |t_{1s}|^2 B_s (-\epsilon_1) +\sum_{s=L,R}  |t_{2s}|^2 B_s (-\epsilon_2-U), \\
	\langle 01 \rvert H_\text{LS} \lvert 01\rangle & = -	\sum_{s=L,R}  |t_{2s}|^2 B_s (-\epsilon_2)+\sum_{s=L,R}  |t_{1s}|^2 B_s (-\epsilon_1-U),\\
	\langle 11 \rvert H_\text{LS} \lvert 11\rangle & =\sum_{j=1,2} 	\sum_{s=L,R}  |t_{js}|^2 B_s (-\epsilon_j-U),\\
	\langle 10 \rvert H_\text{LS} \lvert 01\rangle & = \left(\langle 01 \rvert H_\text{LS} \lvert 10\rangle\right)^* = -\frac{1}{2}\sum_{j=1,2}\sum_{s=L,R} t_{1s}^*t_{2s}\left[ B_s(-\epsilon_j)+  B_s(-\epsilon_j-U)\right].
\end{aligned}
  	\label{app_eq:HLS}
\end{equation}
\end{widetext}

The Lindblad dynamics feature a weak symmetry with respect to the number of electrons $N_\text{PD}$ on the dots. From \cref{app_eq:HQD,app_eq:HLS}, the effective Hamiltonian conserves the number of electrons, $[N_\text{PD},H_\text{eff}]=0$.

\subsection{Jump operators}\label{app:Lindblad_jumps}

To construct the jump operators for  \cref{eq:evolution}, we follow the approach presented in \cite{Kirsanskas2018}. For the model of two parallel quantum dots we  proceed as follows.
\begin{enumerate}
	\item
	Each jump operator is identified with a physical jump process between the quantum dot system and the leads, see~\cref{eq:operator_sys}.  For the parallel dots there are eight jump processes in total.  
	\begin{enumerate}
		\item
		The electron arrives from the left lead to dot 1 (or dot 2), i.e.,  $X^{(1)}$  [or $X^{(2)}$], with  the corresponding tunneling amplitude $t^*_{1L}$ (or $t^*_{2L}$).
		\item
		The electron leaves dot 1 (or dot 2) into the left lead, $X^{(3)}$ [or $X^{(4)}$], with the tunneling amplitude  $t_{1L}$ (or $t_{2L}$).
		\item
		The electron arrives from the right lead to dot 1 (or dot 2), i.e., $X^{(1)}$  [or $X^{(2)}$],  the corresponding tunneling amplitude is $t^*_{1R}$ (or $t^*_{2R}$).
		\item
		The electron leaves dot 1 (or dot 2) into the right lead, $X^{(3)}$ [or $X^{(4)}$], with the tunneling amplitude $t_{1R}$ (or $t_{2R}$).
	\end{enumerate}
	\item
	 The processes for each lead are combined \cite{Kirsanskas2018}, which yields
\begin{equation}
	\begin{aligned}
		\tilde{J}_{+ L} &= t^*_{1L} X^{(1)}+ t^*_{2L} X^{(2)}, \\
		\tilde{J}_{- L} &=  t_{1L} X^{(3)} + t_{2L} X^{(4)}, \\
		\tilde{J}_{+ R} &=  t^*_{1R} X^{(1 )}+ t^*_{2R} X^{(2)},\\
		\tilde{J}_{- R} & =  t_{1R} X^{(3)} + t_{2R} X^{(4)}.
\end{aligned}
\end{equation}
	\item
	Finally,  each jump operator  is reweighted in the eigenbasis of $H_\text{PD}$ of \cref{eq:HQD} according to the energy differences, by the corresponding distribution of the resonant energy in the lead before the electron exchange [cf.~\cref{eq:Fermi}], and the corresponding density of states $\nu$ (assumed constant),
\begin{equation}
	\begin{aligned}
	    &\left(J_{+ s}\right)_{mn} \\& \quad = \left(\tilde{J}_{+ s}\right)_{mn}  \sqrt{2\pi\nu}\sqrt{f_s\left(E_m - E_n\right)},\\
	      &\left(J_{- s}\right)_{mn}  \\& \quad = \left(\tilde{J}_{- s}\right)_{mn}  \sqrt{2\pi\nu}\sqrt{1-f_s\left(E_n - E_m\right)},
\end{aligned}
\end{equation}
where $f_s(\epsilon)$ is the Fermi distribution for energy $\epsilon $ in lead $s=L,R$,
so that 
	\begin{widetext} 
    \begin{equation}
	\begin{aligned}
		J_{+ L} &=\sqrt{2\pi \nu}
		\begin{pmatrix}
			0 & 0 & 0 & 0 \\
			 t_{1L}^* \sqrt{f_L\left(\epsilon_1\right)} & 0 & 0 & 0 \\
			 t_{2L}^*\sqrt{f_L\left(\epsilon_2\right)} & 0 & 0 & 0 \\
			0 & -t_{2L}^* \sqrt{f_L\left(\epsilon_2 + U\right)} &   t_{1L}^*\sqrt{f_L\left(\epsilon_1 + U\right)}& 0
		\end{pmatrix}, \\
		J_{- L} &= \sqrt{2\pi \nu}
		\begin{pmatrix}
			0 & t_{1L} \sqrt{1 - f_L\left(\epsilon_1\right)}  &  t_{2L} \sqrt{1 - f_L\left(\epsilon_2\right)} & 0 \\
			0 & 0 & 0 & - t_{2L}  \sqrt{1 - f_L\left( \epsilon_2 + U\right)}  \\
			0 & 0 & 0 &   t_{1L} \sqrt{1 - f_L\left(\epsilon_1 + U\right)} \\
			0 & 0 & 0 & 0
		\end{pmatrix}, \\
		J_{+ R} &= \sqrt{2\pi \nu}
		\begin{pmatrix}
			0 & 0 & 0 & 0 \\
			 t_{1R}^*  \sqrt{f_R\left(\epsilon_1\right)} & 0 & 0 & 0 \\
			 t_{2R}^* \sqrt{f_R\left(\epsilon_2\right)} & 0 & 0 & 0 \\
			0 & -t_{2R}^*  \sqrt{f_R\left(\epsilon_2 + U\right)} &   t_{1R}^* \sqrt{f_R\left(\epsilon_1 + U\right)}& 0
		\end{pmatrix},  \\
	 	J_{- R}& = \sqrt{2\pi \nu}
		\begin{pmatrix}
			0 & t_{1R}  \sqrt{1 - f_R\left( \epsilon_1\right)}  &  t_{2R} \sqrt{1 - f_R\left(\epsilon_2\right)} & 0 \\
			0 & 0 & 0 & - t_{2R}  \sqrt{1 - f_R\left( \epsilon_2 + U\right)}  \\
			0 & 0 & 0 &  t_{1R} \sqrt{1 - f_R\left(\epsilon_1 + U\right)} \\
			0 & 0 & 0 & 0
		\end{pmatrix},
	\end{aligned}
		\label{app_eq:jump}
	\end{equation}
 \end{widetext}
in the basis of $|00\rangle$, $|10\rangle$, $|01\rangle$ and $|11\rangle$.
\end{enumerate}
The jump operators of ~\cref{app_eq:jump} increase or decrease  the number of electrons only by $1$,
\begin{equation}\label{app_eq:jump_weak}
[N_\text{PD},J_{\alpha s}]=\alpha \,J_{\alpha s},
\end{equation}
where $\alpha=+,-$ and $s=L,R$. In particular, \cref{app_eq:jump_weak} leads to $ [N_\text{PD},J_{\alpha s}^\dagger J_{\alpha s}]=0$, so that the average particle current in \cref{eq:current} is determined only by the components $\rho_\text{PD}(t)$ diagonal in charge. Therefore, the  left and right eigenmatrices of the Liouville operator can be chosen as eigenmatrices of $\mathcal{N}_\text{PD}$, see also \cref{app:perturbation_0}.

\section{Perturbative dynamics and metastability} \label{app:pt_meta}

\subsection{Dynamics with strong symmetry}\label{app:perturbation_0}

Here, we discuss further the dynamics in the presence of the strong swap symmetry,  cf.~Eqs.~\eqref{eq:upb_Heff} and~\eqref{eq:upb_jump}. 

Next to the stationary states in \cref{eq:unp_stationary} and the projections in \cref{eq:unp_proj}, there are two decay modes corresponding to the classical dynamics, 
\begin{equation}
	\begin{aligned}
		R_5^{(0)} &=\lvert 00 \rangle\!\langle 00\lvert -\vert + \rangle\!\langle +\lvert, \\
		R_6 ^{(0)}& = \lvert - \rangle\!\langle -\lvert -\vert 11 \rangle\!\langle 11 \lvert, 
	\end{aligned}
	\label{eq:unp_decay_r}
\end{equation}
and 
\begin{equation}
	\begin{aligned}
		L_5^{(0)} &=\bar{f} (\epsilon) \lvert 00 \rangle\!\langle 00\lvert-[1-\bar{f} (\epsilon)]\vert + \rangle\!\langle +\lvert,\\
		L_6^{(0)} &=\bar{f} (\epsilon+U)\lvert - \rangle\!\langle -\lvert -[1-\bar{f} (\epsilon+U)]\vert 11 \rangle\!\langle 11 \lvert, 
	\end{aligned}
	\label{eq:unp_decay_l}
\end{equation}
with the degenerate pair of eigenvalues given by \cref{eq:lambda56}

There is also a decay of the quantum coherences in the bonding/anti-bonding basis, 
\begin{align}\label{eq:unp_coh_r}
	R_3^{(0)} & = [R_4^{(0)}]^\dagger =\lvert +\rangle\!\langle -\lvert,\\
	\label{eq:unp_coh_l}
	L_3^{(0)} & =  [L_4^{(0)}]^\dagger = \lvert -\rangle\!\langle +\lvert,
\end{align}
with the conjugate pair of eigenvalues in \cref{eq:lambda34}.

\subsection{Perturbation theory for strong symmetry breaking}\label{app:perturbation}

We now investigate dynamics of the Liouville operator $\mathcal{L}$ using the non-Hermitian perturbation theory with respect to perturbations away from dynamics featuring the strong swap symmetry, see~\cref{eq:per_master}.  Those arise due to perturbations in the effective Hamiltonian and the jump operators, cf.~Eqs.~\eqref{eq:upb_Heff} and~\eqref{eq:upb_jump}, when dynamical parameters are changed according to Eqs.~\eqref{eq:per_epsilon} and~\eqref{eq:per_Gamma}.

\subsubsection{Perturbations of Liouvillian}\label{app:perturbation_master}
The resulting perturbations to the Liouville operator caused by perturbations of the dynamical parameters are of all orders. This is due to the fact that the effective Hamiltonian and the jump operators are non-linear functions of the dot energies and the tunneling rates, see \cref{app:Lindblad}.
In particular, we have that the first-order perturbation of the Liouvillian [cf. Eqs.~\eqref{eq:evolution} and~\eqref{eq:per_master}] is given by
\begin{widetext}
\begin{align}
		\mathcal{L}^{(1)}(\rho_\text{PD}) &=		- i \left[H_\text{eff}^{(1)}, \rho_\text{PD}\right]\notag\\
		& + \sum_{\substack{\alpha = +,-\\ s = L,R}} \left\{ J_{\alpha s}^{(1)} \rho_\text{PD} \left[J_{\alpha s} ^{(0)}\right]^\dagger+J_{\alpha s}^{(0)} \rho_\text{PD} \left[J_{\alpha s}^{(1)}\right]^\dagger\right\} 
		- \frac{1}{2}\sum_{\substack{\alpha = +,-\\ s = L,R}}\left\{ \rho_\text{PD}, \left[J_{\alpha s}^{(1)}\right]^\dagger J_{\alpha s}^{(0)}+\left[J_{\alpha s}^{(0)}\right]^\dagger J_{\alpha s}^{(1)}\right\},
		\label{eq:per_master_1}
	\end{align}
which stems from the first-order perturbations to the effective Hamiltonian in \cref{eq:upb_Heff}
and the jump operators in \cref{eq:upb_jump}. 
Similarly, the second-order perturbation of $\mathcal{L}$ is given by
\begin{equation}\label{eq:per_master_2}
	\begin{aligned}
		\mathcal{L}^{(2)}(\rho_\text{PD})= &		- i \left[H_\text{eff}^{(2)}, \rho_\text{PD}\right]
		 + \sum_{\substack{\alpha = +,-\\ s = L,R}} \left\{ J_{\alpha s}^{(2)} \rho_\text{PD} \left[J_{\alpha s} ^{(0)}\right]^\dagger+J_{\alpha s}^{(0)} \rho_\text{PD} \left[J_{\alpha s}^{(2)}\right]^\dagger\right\}\\
		&- \frac{1}{2}\sum_{\substack{\alpha = +,-\\ s = L,R}}\left\{ \rho_\text{PD}, \left[J_{\alpha s}^{(2)}\right]^\dagger J_{\alpha s}^{(0)}+\left[J_{\alpha s}^{(0)}\right]^\dagger J_{\alpha s}^{(2)}\right\},
		 + \sum_{\substack{\alpha = +,-\\ s = L,R}} \left\{ J_{\alpha s}^{(1)} \rho_\text{PD} \left[J_{\alpha s}^{(1)}\right]^\dagger - \frac{1}{2}\left\{ \rho_\text{PD}, \left[J_{\alpha s}^{(1)}\right]^\dagger J_{\alpha s}^{(1)}\right\}\right\},
	\end{aligned}
\end{equation}
where both the first- and second-order perturbations to the effective Hamiltonian and the jump operators contribute. We will denote the contribution from the first-order perturbations only as
\begin{equation}\label{eq:per_master_2'}
	\begin{aligned}
		&\mathcal{L}^{(2)'}(\rho_\text{PD})=	  \sum_{\substack{\alpha = +,-\\ s = L,R}} \left\{ J_{\alpha s}^{(1)} \rho_\text{PD} \left[J_{\alpha s}^{(1)}\right]^\dagger - \frac{1}{2}\left\{ \rho_\text{PD}, \left[J_{\alpha s}^{(1)}\right]^\dagger J_{\alpha s}^{(1)}\right\}\right\},
	\end{aligned}
\end{equation}
Below, we only give first-order perturbations to the effective Hamiltonian and the jump operators, as they fully determine the leading second-order corrections  to the long-time dynamics, via $\mathcal{L}^{(1)}$ and $\mathcal{L}^{(2)'}$, which is argued in \cref{app:perturbation_Leff}.\\

For the dot Hamiltonian of \cref{eq:HQD} the perturbation is linear in $\delta \epsilon$ 
\begin{equation}\label{app_eq:HQD_pb}
	\begin{aligned}
	&\delta H_\text{PD}=H_\text{PD}-H^{(0)}_\text{PD}=H^{(1)}_\text{PD}= \delta \epsilon \left( |+\rangle\!\langle-|+|-\rangle\!\langle+|\right),
\end{aligned}
\end{equation}
see also \cref{eq:per_epsilon}.

For the Lamb shift Hamiltonian, we have, up to the second order in $\delta \epsilon$ and $\delta \Gamma$ [cf.\,\cref{app_eq:HLS} and see  Eqs.~\eqref{eq:per_epsilon}-\eqref{eq:per_Gamma}]
\begin{equation}	\label{app_eq:HLS_pb}
		\begin{aligned}
			\delta H_\text{LS}=&H_\text{LS}-H^{(0)}_\text{LS}=H^{(1)}_\text{LS}+...\\
		=&-\frac{\Gamma }{\pi} \left\{\delta \epsilon \left[ \bar{B}'(-\epsilon)+ \bar{B}'(-\epsilon-U)\right] +\frac{\delta \Gamma}{\Gamma}\left[ \bar{B}^-(-\epsilon)+ \bar{B}^-(-\epsilon-U)\right] \right\} \left(|+\rangle\!\langle -|+ |-\rangle\!\langle +|\right)+...,
\end{aligned}
\end{equation}
where $\bar{B}'(-\epsilon)=\partial_\epsilon\bar{B}(-\epsilon)$ and
 $\bar{B}^-(-\epsilon)= [ B_L(-\epsilon)-  B_R(-\epsilon)]/2 $.\\

For the jump operators, we have up to the second order [cf.\,\cref{app_eq:jump} and see  Eqs.~\eqref{eq:per_epsilon}-\eqref{eq:per_Gamma}]
    \begin{equation}	\label{app_eq:jump_pb}
  	\begin{aligned}
  		\delta J_{+L}&=J_{+ L}-J_{+ L}^{(0)}=J_{+ L}^{(1)}+... \\
  		&= \sqrt{\frac{\Gamma}{2}} \sqrt{f_L\left(\epsilon\right)} \left[ \delta \epsilon\frac{f_L'\left(\epsilon\right)}{f_L\left(\epsilon\right)}-\frac{\delta \Gamma}{\Gamma} \right]|-\rangle\!\langle00| -\sqrt{\frac{\Gamma}{2}} \sqrt{f_L\left(\epsilon+U\right)} \left[ \delta \epsilon\frac{f_L'\left(\epsilon+U\right)}{f_L\left(\epsilon+U\right)}-\frac{\delta \Gamma}{\Gamma} \right]|11\rangle\!\langle+| +...,\\
  		\delta J_{-L}&=J_{- L}-J_{- L}^{(0)} =J_{- L}^{(1)}+...\\
  		&=  -\sqrt{\frac{\Gamma}{2}}\sqrt{1-f_L\left(\epsilon\right)}\left[\delta \epsilon\frac{f_L'\left(\epsilon\right)}{1-f_L\left(\epsilon\right)} +\frac{\delta \Gamma}{\Gamma}\right] |00\rangle\!\langle-| +\sqrt{\frac{\Gamma}{2}}\sqrt{1-f_L\left(\epsilon+U\right)}\left[\delta \epsilon\frac{f_L'\left(\epsilon+U\right)}{1-f_L\left(\epsilon+U\right)} +\frac{\delta \Gamma}{\Gamma}\right]|+\rangle\!\langle11| +...\\
  		  		\delta J_{+R}&=J_{+ R}-J_{+ R}^{(0)}=J_{+ R}^{(1)}+... \\
  		&= \sqrt{\frac{\Gamma}{2}} \sqrt{f_R\left(\epsilon\right)} \left[ \delta \epsilon\frac{f_R'\left(\epsilon\right)}{f_R\left(\epsilon\right)}+\frac{\delta \Gamma}{\Gamma} \right]|-\rangle\!\langle00| -\sqrt{\frac{\Gamma}{2}}\sqrt{f_R\left(\epsilon+U\right)} \left[ \delta \epsilon\frac{f_R'\left(\epsilon+U\right)}{f_R\left(\epsilon+U\right)}+\frac{\delta \Gamma}{\Gamma} \right]|11\rangle\!\langle+| +...,\\
  		\delta J_{-s}&=J_{- s}-J_{- s}^{(0)} =J_{- s}^{(1)}+...\\
  		&= - \sqrt{\frac{\Gamma}{2}}\sqrt{1-f_R\left(\epsilon\right)}\left[\delta \epsilon\frac{f_R'\left(\epsilon\right)}{1-f_R\left(\epsilon\right)} -\frac{\delta \Gamma}{\Gamma}\right] |00\rangle\!\langle-| +\sqrt{\frac{\Gamma}{2}}\sqrt{1-f_R\left(\epsilon+U\right)}\left[\delta \epsilon\frac{f_R'\left(\epsilon+U\right)}{1-f_R\left(\epsilon+U\right)} -\frac{\delta \Gamma}{\Gamma}\right]|+\rangle\!\langle11|+...,%\\
  	\end{aligned}
  \end{equation}
  where $f_s'(\epsilon)=\partial_\epsilon f_s(\epsilon)$.\\
\end{widetext}

We note that the effective Hamiltonian and the jump operators feature symmetry-breaking perturbations only in the first order. In fact, it can be shown that symmetry-breaking perturbations of those operators appear in odd orders, while symmetry-preserving perturbations appear in even orders. This is a consequence of the fact that choosing perturbations with the opposite signs in Eqs.~\eqref{eq:per_epsilon} and~\eqref{eq:per_Gamma}, directly corresponds to the dynamics with the dots swapped. Under this transformation,  $|-\rangle$ is replaced by $-|-\rangle$ and $|11\rangle$ by  $-|11\rangle$ in the bonding/anti-bonding basis.  Since the simultaneous change of all perturbation signs changes the sign of odd order corrections, those must correspond to the  symmetry-breaking contributions, while even orders must be accompanied only by symmetry-preserving contributions.

\subsubsection{Perturbative corrections to dynamics}\label{app:perturbation_Leff}

As the dynamics, no matter the size of perturbation, preserve the weak symmetry with respect to $N_\text{PD}$, the only unperturbed modes that contribute to the perturbed dynamics of$\rho_\text{PD}$, as considered in the main text, are the modes diagonal in charge [see \cref{sec:weak,app:Lindblad}].

We consider the perturbation theory for the reduced dynamics of the first two eigenmodes of the dynamics, $\mathcal{L}\mathcal{P}$, where $\mathcal{P}(\rho_\text{PD})=\rho_\text{ss}+ \mathrm{Tr}(L_2\rho_\text{PD}) R_2$ [cf. Eqs.~\eqref{eq:evolution3} and~\eqref{eq:evolution_meta}]. 
We have that $\mathcal{P}=\mathcal{P}^{(0)}+\mathcal{P}^{(1)}+...$ where $\mathcal{P}^{(0)}$ is the projection on the zero-eigenspace of the unperturbed dynamics $\mathcal{L}^{(0)}$,
\begin{align}
	\mathcal{P}^{(0)}(\rho_\text{PD})=\sum_{i=1,2}\rho_{i}^\text{ss} \mathrm{Tr}(P_i\rho_\text{PD}).
	\label{eq:P0}
\end{align}

The first-order corrections to the reduced dynamics are always within that subspace and are formally given by $\mathcal{P}^{(0)}\mathcal{L}^{(1)}\mathcal{P}^{(0)}$ \cite{Kato}. We now show that these corrections are zero for the dynamics considered in this work. This can be seen as the consequence of the classicality of the zero-eigenspace of the unperturbed dynamics (cf. Ref.\,\cite{Macieszczak2021}).  

In the first order, the perturbations of the effective Hamiltonian and the jump operators are symmetry-breaking and give rise to the first-order corrections to the Liouvillian $\mathcal{L}^{(1)}$, given in \cref{eq:per_master_1}, which break the strong symmetry.  That is, $\mathcal{L}^{(1)}(\rho_i^\text{ss})$ is a linear combination of coherences $|+\rangle\!\langle-|$ and $|-\rangle\!\langle+|$. Since the coherences decay to $0$ under the unperturbed dynamics, $\mathrm{Tr}(P_i |+\rangle\!\langle-|)=0=\mathrm{Tr}(P_i |-\rangle\!\langle+|)$, the first-order corrections vanish, $\mathcal{P}^{(0)}\mathcal{L}^{(1)}\mathcal{P}^{(0)}=0$. \\

For $\mathcal{P}^{(0)}\mathcal{L}^{(1)}\mathcal{P}^{(0)}=0$, the second-order corrections to the reduced dynamics,
are found within the zero-subspace of $\mathcal{L}^{(0)}$ and formally given by  
$\mathcal{P}^{(0)}\mathcal{L}^{(2)}\mathcal{P}^{(0)}-\mathcal{P}^{(0)}\mathcal{L}^{(1)}\mathcal{R}^{(0)}\mathcal{L}^{(1)}\mathcal{P}^{(0)}$\cite{Kato},
where $\mathcal{R}^{(0)}$ is the reduced resolvent of $\mathcal{L}^{(0)}$ at 0, so that $\mathcal{R}^{(0)}\mathcal{L}^{(0)}~=\mathcal{L}^{(0)}\mathcal{R}^{(0)}=\mathcal{I}-\mathcal{P}^{(0)}$, with the identity map $\mathcal{I}(\rho_{PD})~=\rho_{PD}$. In terms of the eigenmatrices of $\mathcal{L}^{(0)}$ we can write [\cref{eq:unp_decay_r,eq:unp_coh_r,eq:lambda34,eq:lambda56}]
\begin{align}\label{eq:R0}
	\mathcal{R}^{(0)}(\rho_\text{PD})=\sum_{i=3}^6 \frac{1}{\lambda_i} R_i^{(0)} \mathrm{Tr}\!\left[L_i^{(0)}\!\rho_\text{PD}\right]. 
\end{align}
We now show that the contribution to the second-order dynamics stem only from the first-order corrections to the effective Hamiltonian and the jump operators.

Indeed, let us note that $\mathcal{L}^{(2)}-\mathcal{L}^{(2)'}$ is of an analogous form to $\mathcal{L}^{(1)}$ but with the first-order perturbations $H_\text{eff}^{(1)}$ and $J_{\alpha s}^{(1)}$ replaced by the second-order perturbations $H_\text{eff}^{(2)}$ and $J_{\alpha s}^{(2)}$ [cf. Eqs.~\eqref{eq:per_master_1}-\eqref{eq:per_master_2'}]. Those perturbations are symmetry-preserving, $[\mathcal{L}^{(2)}-\mathcal{L}^{(2)'}]^{\dagger}(P_i) =0$ for $i=1,2$, and the trace within the support of $\rho_i^\text{ss}$ is preserved by $\mathcal{L}^{(2)}-\mathcal{L}^{(2)'}$, which leads to $\mathcal{P}^{(0)}[\mathcal{L}^{(2)}-\mathcal{L}^{(2)'}]~=0$. Thus, the second-order corrections to the effective Hamiltonian and the jump operators do not contribute to second-order corrections in the reduced dynamics.

We conclude that the second-order corrections to the reduced dynamics are given by  $\mathcal{P}^{(0)}\mathcal{L}^{(2)'}\mathcal{P}^{(0)}-\mathcal{P}^{(0)}\mathcal{L}^{(1)}\mathcal{R}^{(0)}\mathcal{L}^{(1)}\mathcal{P}^{(0)}$.
	We now use this results to calculate the decay rates in \cref{eq:eff_dynamics}. In the operator basis of $\lvert +\rangle\!\langle -\lvert, \lvert -\rangle\!\langle +\lvert , \lvert 00\rangle\!\langle 00\vert ,\lvert +\rangle\!\langle +\lvert , \lvert -\rangle\!\langle -\lvert , \lvert 11\rangle\!\langle 11\lvert $ (which we index by $\text{I}$, $\text{II}$, $\text{III}$, $\text{IV}$, $\text{V}$, $\text{VI}$, respectively), the first-order  perturbation $\mathcal{L}^{(1)}$ of the Liouvillian has the following structure 
	\begin{widetext}
	\begin{align}
		\doublehat{\mathcal{L}}^{(1)} = 
		\left(
		\begin{array}{cccccc}
			0 & 0 &   \doublehat{\mathcal{L}}^{(1)}_\text{I,III} & \doublehat{\mathcal{L}}^{(1)}_\text{I,IV} & \doublehat{\mathcal{L}}^{(1)}_\text{I,V}&  \doublehat{\mathcal{L}}^{(1)}_\text{I,VI}\\
			0 & 0 &  \left[\doublehat{\mathcal{L}}^{(1)}_\text{I,III}\right]^* &  \left[\doublehat{\mathcal{L}}^{(1)}_\text{I,IV}\right]^* &  \left[\doublehat{\mathcal{L}}^{(1)}_\text{I,V}\right]^*  &  \left[\doublehat{\mathcal{L}}^{(1)}_\text{I,VI}\right]^*\\
			\doublehat{\mathcal{L}}^{(1)}_\text{III,I} &  \left[\doublehat{\mathcal{L}}^{(1)}_\text{III,I} \right]^*& 0 & 0 & 0 & 0\\
		\left[\doublehat{\mathcal{L}}^{(1)}_\text{I,V}\right]^*  &  \doublehat{\mathcal{L}}^{(1)}_\text{I,V} & 0 & 0 & 0 & 0\\
			 \left[\doublehat{\mathcal{L}}^{(1)}_\text{I,IV}\right]^*&  \doublehat{\mathcal{L}}^{(1)}_\text{I,IV} & 0 & 0 &  0 & 0 \\
			\doublehat{\mathcal{L}}^{(1)}_\text{VI,I} &  \left[\doublehat{\mathcal{L}}^{(1)}_\text{VI,I}\right]^* & 0 & 0 & 0 &0 
		\end{array}
		\right).
		\label{eq:matrix_L1}
	\end{align}
	Here, the complex conjugation relations between the first and second columns, and between the first and  second rows, follow from the fact that $\mathcal{L}^{(1)}$ is Hermiticity preserving. We also have $\doublehat{\mathcal{L}}^{(1)}_\text{I,IV}=\doublehat{\mathcal{L}}^{(1)}_\text{V,II}$,
	$\doublehat{\mathcal{L}}^{(1)}_\text{I,V}=\doublehat{\mathcal{L}}^{(1)}_\text{IV,II}$, 
	$\doublehat{\mathcal{L}}^{(1)}_\text{II,IV}=\doublehat{\mathcal{L}}^{(1)}_\text{V,I}$, and
	$\doublehat{\mathcal{L}}^{(1)}_\text{II,V}=\doublehat{\mathcal{L}}^{(1)}_\text{IV,I}$, 
	as those contributions arise only due to the effective Hamiltonian. 
	
	The structure of the second-order perturbation $\mathcal{L}^{(2)'}$ due to the first-order  perturbations of the jump operators,
	\begin{align}
		\doublehat{\mathcal{L}}^{(2)'} = 
		\left(
		\begin{array}{cccccc}
			\left[\doublehat{\mathcal{L}}^{(2)'}\right]_\text{I,I} & 0 & 0 & 0 & 0 & 0\\
			0 &  \left[\doublehat{\mathcal{L}}^{(2)'}\right]_\text{I,I}^* & 0 & 0 & 0 & 0\\
			0 & 0 & -\doublehat{\mathcal{L}}^{(2)}_\text{V,III} & 0 & \doublehat{\mathcal{L}}^{(2)}_\text{III,V} & 0\\
			0 & 0 & 0 & -\doublehat{\mathcal{L}}^{(2)}_\text{VI,IV} & 0 & \doublehat{\mathcal{L}}^{(2)}_\text{IV,VI} \\
			0 & 0 & \doublehat{\mathcal{L}}^{(2)}_\text{V,III} & 0 &  -\doublehat{\mathcal{L}}^{(2)}_\text{III,V} & 0\\
			0 & 0 & 0 & \doublehat{\mathcal{L}}^{(2)}_\text{VI,IV} &   0 & -\doublehat{\mathcal{L}}^{(2)}_\text{IV,VI}
		\end{array}
		\right),
		\label{eq:matrix_L2}
	\end{align}
	can be understood as corresponding to the strong symmetry with $|00\rangle\!\langle 00|-|+\rangle\!\langle+|+|-\rangle\!\langle-|-|11\rangle\!\langle11|$, with which the first-order perturbations of the jump operators  commute. We then use the trace-preservation of $\mathcal{L}^{(2)'}$ to connect the diagonal terms to the off-diagonal ones, and its Hermiticity-preservation to note that $[\doublehat{\mathcal{L}}^{(2)'}]_\text{II,II}=[\doublehat{\mathcal{L}}^{(2)'}]_\text{I,I}^*$.

Therefore, the decay rates in \cref{eq:eff_dynamics} are
	\begin{equation}
		\begin{aligned}
			\gamma_1 &=  \left[1 - \bar{f}(\epsilon) \right] \doublehat{\mathcal{L}}^{(2)}_\text{V,III} +  \bar{f}(\epsilon)\, \doublehat{\mathcal{L}}^{(2)}_\text{VI,IV}\\ 
		& + \frac{1}{2\Gamma} \left( \frac{\left[1 - \bar{f}(\epsilon) \right]\doublehat{\mathcal{L}}^{(1)}_\text{I,III} +  \bar{f}(\epsilon) \doublehat{\mathcal{L}}^{(1)}_\text{I,IV} }{ 1-\bar{f}(\epsilon)+\bar{f}(\epsilon+U)	-\frac{i}{\pi} \left[\bar{B}(-\epsilon) + \bar{B}(-\epsilon - U)\right]} \left\{\doublehat{\mathcal{L}}^{(1)}_\text{III,I} + \left[\doublehat{\mathcal{L}}^{(1)}_\text{I,V}\right]^*\right\}+\text{h.c.}\right),\\
			\gamma_2 &=  \left[1 - \bar{f}(\epsilon+U) \right] \doublehat{\mathcal{L}}^{(2)}_\text{III,V} +  \bar{f}(\epsilon+U)\doublehat{\mathcal{L}}^{(2)}_\text{IV,VI} \\
			& + \frac{1}{2\Gamma} \left( \frac{\left[1 - \bar{f}(\epsilon+U) \right]\doublehat{\mathcal{L}}^{(1)}_\text{I,V} +  \bar{f}(\epsilon+U) \doublehat{\mathcal{L}}^{(1)}_\text{I,VI} }{ 1-\bar{f}(\epsilon)+\bar{f}(\epsilon+U)	-\frac{i}{\pi} \left[\bar{B}(-\epsilon) + \bar{B}(-\epsilon - U)\right]} \left\{\left[\doublehat{\mathcal{L}}^{(1)}_\text{I,IV}\right]^* + \doublehat{\mathcal{L}}^{(1)}_\text{VI,I}\right\}+\text{h.c.}\right),
		\end{aligned}
		\label{eq:exp_gamma12}
	\end{equation}
where
\begin{equation}
	\begin{aligned}
	\doublehat{\mathcal{L}}^{(1)}_\text{I,III} &=   \Gamma \left\{ f_L(\epsilon)\left[ \delta \epsilon\frac{f_L'\left(\epsilon\right)}{f_L\left(\epsilon\right)}-\frac{\delta \Gamma}{\Gamma} \right]+ f_R(\epsilon)\left[ \delta \epsilon\frac{f_R'\left(\epsilon\right)}{f_R\left(\epsilon\right)}+\frac{\delta \Gamma}{\Gamma} \right]\right\},\\
	\doublehat{\mathcal{L}}^{(1)}_\text{I,IV} &= i \left(\delta \epsilon-\frac{\Gamma }{\pi} \left\{\delta \epsilon \left[ \bar{B}'(-\epsilon)+ \bar{B}'(-\epsilon-U)\right] +\frac{\delta \Gamma}{\Gamma} \left[ \bar{B}^-(-\epsilon)+ \bar{B}^-(-\epsilon-U)\right] \right\} \right)\\
	&-   \frac{\Gamma}{2} \left\{ f_L\left(\epsilon+U\right) \left[ \delta \epsilon\frac{f_L'\left(\epsilon+U\right)}{f_L\left(\epsilon+U\right)}-\frac{\delta \Gamma}{\Gamma} \right]+f_R\left(\epsilon+U\right) \left[ \delta \epsilon\frac{f_R'\left(\epsilon+U\right)}{f_R\left(\epsilon+U\right)}+\frac{\delta \Gamma}{\Gamma} \right]\right\}\\
	&-   \frac{\Gamma}{2} \left\{ \left[1-f_L\left(\epsilon\right)\right] \left[ \delta \epsilon\frac{f_L'\left(\epsilon\right)}{1-f_L\left(\epsilon\right)}+\frac{\delta \Gamma}{\Gamma} \right]+ \left[1-f_R\left(\epsilon\right)\right] \left[ \delta \epsilon\frac{f_R'\left(\epsilon\right)}{1-f_R\left(\epsilon \right)}-\frac{\delta \Gamma}{\Gamma} \right]\right\}=\left[	\doublehat{\mathcal{L}}^{(1)}_\text{I,V} \right]^*,\\
			\doublehat{\mathcal{L}}^{(1)}_\text{I,VI} &=   \Gamma \left\{ \left[1-f_L(\epsilon+U)\right]\left[ \delta \epsilon\frac{f_L'\left(\epsilon+U\right)}{1-f_L\left(\epsilon+U\right)}+\frac{\delta \Gamma}{\Gamma} \right]+ \left[1-f_R(\epsilon+U)\right]\left[ \delta \epsilon\frac{f_R'\left(\epsilon+U\right)}{1-f_R\left(\epsilon+U\right)}-\frac{\delta \Gamma}{\Gamma} \right]\right\},\\
				\doublehat{\mathcal{L}}^{(1)}_\text{III,I} &=   -\Gamma \left\{ \left[1-f_L(\epsilon)\right]\left[ \delta \epsilon\frac{f_L'\left(\epsilon\right)}{1-f_L\left(\epsilon\right)}+\frac{\delta \Gamma}{\Gamma} \right]+ \left[1-f_R(\epsilon)\right]\left[ \delta \epsilon\frac{f_R'\left(\epsilon\right)}{1-f_R\left(\epsilon\right)}-\frac{\delta \Gamma}{\Gamma} \right]\right\},\\
							\doublehat{\mathcal{L}}^{(1)}_\text{VI,I} &= -  \Gamma \left\{ f_L(\epsilon+U)\left[ \delta \epsilon\frac{f_L'\left(\epsilon+U\right)}{f_L\left(\epsilon+U\right)}-\frac{\delta \Gamma}{\Gamma} \right]+ f_R(\epsilon+U)\left[ \delta \epsilon\frac{f_R'\left(\epsilon+U\right)}{f_R\left(\epsilon+U\right)}+\frac{\delta \Gamma}{\Gamma} \right]\right\},
	\end{aligned}
\end{equation}
and
\begin{equation}
	\begin{aligned}
		\doublehat{\mathcal{L}}^{(2)}_\text{V,III} &= \frac{ \Gamma}{2} \left\{ f_L(\epsilon)\left[ \delta \epsilon\frac{f_L'\left(\epsilon\right)}{f_L\left(\epsilon\right)}-\frac{\delta \Gamma}{\Gamma} \right]^2+ f_R(\epsilon)\left[ \delta \epsilon\frac{f_R'\left(\epsilon\right)}{f_R\left(\epsilon\right)}+\frac{\delta \Gamma}{\Gamma} \right]^2\right\},\\
		\doublehat{\mathcal{L}}^{(2)}_\text{VI,IV} &=\frac{ \Gamma}{2}  \left\{ f_L(\epsilon+U)\left[ \delta \epsilon\frac{f_L'\left(\epsilon+U\right)}{f_L\left(\epsilon+U\right)}-\frac{\delta \Gamma}{\Gamma} \right]^2+ f_R(\epsilon+U)\left[ \delta \epsilon\frac{f_R'\left(\epsilon+U\right)}{f_R\left(\epsilon+U\right)}+\frac{\delta \Gamma}{\Gamma} \right]^2\right\},\\
		\doublehat{\mathcal{L}}^{(2)}_\text{III,V} &=  \frac{ \Gamma}{2}  \left\{ \left[1-f_L(\epsilon)\right]\left[ \delta \epsilon\frac{f_L'\left(\epsilon\right)}{1-f_L\left(\epsilon\right)}+\frac{\delta \Gamma}{\Gamma} \right]^2+ \left[1-f_R(\epsilon)\right]\left[ \delta \epsilon\frac{f_R'\left(\epsilon\right)}{1-f_R\left(\epsilon\right)}-\frac{\delta \Gamma}{\Gamma} \right]^2\right\},\\
		\doublehat{\mathcal{L}}^{(2)}_\text{IV,VI} &=\frac{ \Gamma}{2} \left\{ \left[1-f_L(\epsilon+U)\right]\left[ \delta \epsilon\frac{f_L'\left(\epsilon+U\right)}{1-f_L\left(\epsilon+U\right)}+\frac{\delta \Gamma}{\Gamma} \right]^2+ \left[1-f_R(\epsilon+U)\right]\left[ \delta \epsilon\frac{f_R'\left(\epsilon+U\right)}{1-f_R\left(\epsilon+U\right)}-\frac{\delta \Gamma}{\Gamma} \right]^2\right\}.		
	\end{aligned}
\end{equation}
using \cref{app_eq:HQD_pb,app_eq:HLS_pb,app_eq:jump_pb}.
\end{widetext}

Third-order corrections to the reduced dynamics projected on the zero subspace of $\mathcal{L}^{(0)}$ in general contribute to the first-order perturbations of the stationary state in \cref{eq:per_ss} and the eigenmatrices in \cref{eq:per_lambda2} corresponding to the second eigenvalue. 
But here the third-order corrections (see Supplemental Material of Ref.\,\cite{Macieszczak2016} and \cite{Kato}) are projected to 0 as, when acting on $\rho_1^\text{ss}$ and $\rho_2^\text{ss}$, they give rise to linear combinations of $|+\rangle\langle-|$ and $|-\rangle\langle+|$. In particular, we have, $\lambda_2^{(3)}=0$. These results can be seen to hold for all odd-order corrections, as the opposite sign of the perturbations in \cref{eq:per_epsilon,eq:per_Gamma} corresponds to the dynamics with the dots swapped, which however leaves $\rho_1^\text{ss}$ and $\rho_2^\text{ss}$ unchanged.

\subsubsection{Perturbative corrections to metastable and stationary states}\label{app:perturbation_rhoss}

 Beyond the zero-subspace spanned by the stationary states of $\mathcal{L}^{(0)}$ in \cref{eq:unp_stationary}, the first-order corrections to the projection $\mathcal{P}$ on the stationary state and the second eigenmode are given by $\mathcal{P}^{(1)}=-\mathcal{R}^{(0)} \mathcal{L}^{(1)} \mathcal{P}^{(0)}$  \cite{Kato}. These corrections determine the first-order corrections to the metastable phases as $-\mathcal{R}^{(0)} \mathcal{L}^{(1)} (\rho^\text{ss}_1)$ and $-\mathcal{R}^{(0)} \mathcal{L}^{(1)} (\rho^\text{ss}_2)$;  see Supplemental Material of Ref.\,\cite{Macieszczak2016}.

 We now assess the first-order corrections to the unique stationary state for  $\mathcal{L}$, cf.\,\cref{eq:per_master}. Again, due to the weak symmetry with respect to $N_\text{PD}$, the only unperturbed modes that contribute are the symmetric ones, cf.~Appendix~\ref{app:perturbation_0}. 
The first-order corrections to the stationary state in \cref{eq:per_ss} are given by 
\begin{align}
		\rho_\text{PD}^{\text{ss}(1)}=& -\mathcal{R}^{(0)} \mathcal{L}^{(1)} \big[\rho_\text{PD}^{\text{ss}(0)} \big].
\end{align}
Here, the projected third-order corrections to the reduced dynamics should also contribute (cf. Supplemental Material of Ref.\,\cite{Macieszczak2016} and \cite{Kato}),  but they vanish for the considered perturbations, as we explained above.

\subsubsection{Perturbative corrections to initial dynamics} \label{app:perturbation_initial}

Generally, the dynamics taking place before the metastable regime can be analyzed in terms of the perturbative corrections to the remaining fast eigenmodes. The leading corrections for the eigenvalues are of second order, but for the eigematrices  of first order.

The eigenvalues corresponding to the coherence decay in \cref{eq:lambda34} will acquire  second-order corrections as in the first order
$\lambda_{3,4}^{(1)}= \mathrm{Tr}[L_{3,4}^{(0)} \mathcal{L}^{(1)} R_{3,4}^{(0)}]=0$, as the symmetry-breaking perturbations of the effective Hamiltonian and jumps cannot contribute here.
The first-order corrections to the corresponding eigenmatrices in \cref{eq:unp_coh_r} and~\eqref{eq:unp_coh_l} will be present in general.

The degeneracy of the classical decay of eigenvalues  in \cref{eq:lambda56}   will only be lifted in the second order, with the first-order corrections being zero (cf.~\cref{fig:spectrum}). The first order corrections to the corresponding matrices in Eqs.~\eqref{eq:unp_decay_r} and~\eqref{eq:unp_decay_l} will be present.

\subsection{Metastable phases and long-time dynamics\\ -- all orders} \label{app:num_metastability}

Formally, the long-time dynamics in \cref{eq:evolution_meta} is a projection onto the subspace of the first two eigenmatrices of the Liouvillian. Here, we review the construction first introduced in  \cite{Macieszczak2016} that allows for considering it in a physical basis, and thus considering \cref{eq:eff_dynamics} not only up the second, but to all orders, as in \cref{fig:metastable_current}(b).

\subsubsection{Metastable phases}\label{app:num_metastability_meta}
To explain how to construct the metastable states in terms of the first two eigenmodes of the dynamics, as used in \cref{fig:metastable_current}(b),  we follow \cite{Macieszczak2016, Rose2016}.

The metastable manifold is spanned by two extreme metastable states  
\begin{equation}
\begin{aligned}
	\tilde{\rho}_1  & = \rho_\text{PD}^\text{ss} + c_2^\text{max} R_2, \\
	\tilde{\rho}_2 & = \rho_\text{PD}^\text{ss} + c_2^\text{min}  R_2.
	\label{eq:eMS}
\end{aligned}
\end{equation}
The coefficients $c_2^\text{min}$ and $c_2^\text{max}$ are the smallest and largest eigenvalues of the left eigenmatrix $L_2$. Using the results of \cref{app:perturbation}, up to the first-order corrections, we obtain $\tilde{\rho}_1=\rho^\text{ss}_1+...$ and $\tilde{\rho}_2=\rho^\text{ss}_2+...$. 

The approximation in \cref{eq:evolution_meta} for any state during the metastable regime corresponds to the projection $\mathcal{P}$ on the stationary state and the second eigenmode, and can be equivalently expressed as a linear combination
\begin{align}
	\mathcal{P}[\rho_\text{PD}(t)] = \tilde{p}_1(t) \tilde{\rho}_1 + \tilde{p}_2(t) \tilde{\rho_2},
	\label{eq:mixture}
\end{align}
with  $\tilde{p}_i(t)$ defined via the observables
\begin{equation}
\begin{aligned}
	\tilde{P}_1 & = \left(L_2 - c_2^\text{min} \mathbb{1} \right)/\Delta c_2, \\
	\tilde{P}_2 & =  \left(-L_2 + c_2^\text{max} \mathbb{1} \right)/\Delta c_2,
\end{aligned}
\label{eq:Pobservable}
\end{equation}
with $\Delta c_2 = c_2^\text{max} -c_2^\text{min}$ as $\tilde{p}_{i}(t) = \text{Tr}[\tilde{P}_{i} \rho_\text{PD}(t)]$ for $i = 1,2$. The metastable phases $\tilde{\rho}_i$ defined above feature trace $1$, but are in general not positive and thus are not described by density matrices. In contrast, $\tilde{p_i}(t)$ always correspond to probabilities \cite{Macieszczak2016, Rose2016}. Up to the first-order corrections, we have $\tilde{P}_{i}=P_i+...$, so that  $\tilde{p_i}(0)=p_i(0)+...$.

\subsubsection{Long-time dynamics}\label{app:num_metastability_Leff}
The time evolution of $\mathcal{P}[\rho_\text{PD}(t)]$ corresponds to the evolution of the probabilities in \cref{eq:mixture} governed by the generator
\begin{equation}
	\frac{d}{dt} 
	\left[\begin{array}{cc}
		\tilde{p}_1(t) \\
		\tilde{p}_2(t)
	\end{array}\right]
	=
	-\frac{\lambda_2}{\Delta c_2}
	\begin{pmatrix}
		-c_2^\text{max} & -c_2^\text{min} \\
		\phantom{-}c_2^\text{max} &	\phantom{-}c_2^\text{min}
	\end{pmatrix}
	\left[\begin{array}{cc}
		\tilde{p}_1(t) \\
		\tilde{p}_2(t)
	\end{array}\right],
	\label{eq:dynamics_eff}
\end{equation} 	
which represents classical stochastic dynamics \cite{Macieszczak2016, Rose2016}. In the second order, the generator in \cref{eq:dynamics_eff} coincides with the one in \cref{eq:eff_dynamics}, cf. Supplemental Material in Ref.\,\cite{Macieszczak2021}.

\subsubsection{Stationary state}\label{app:num_metastability_rhoss}

In terms of the two metastable phases in \cref{eq:eMS} the stationary state decomposes as 
\begin{align}
\rho_\text{PD}^\text{ss} = \tilde{p}_1^\text{ss} \tilde{\rho}_1 +  \tilde{p}_2^\text{ss} \tilde{\rho_2},
	\label{eq:mixture_ss}
\end{align}
where $\tilde{p}_1^\text{ss} $ and $\tilde{p}_2^\text{ss} $ are the stationary probabilities for the classical dynamics in \cref{eq:dynamics_eff} [cf.\,\cref{eq:mixture}], which in the zero-th order equal the stationary probabilities of \cref{eq:eff_dynamics}. Thus, \cref{eq:mixture_ss} in the zero-th order corresponds to \cref{eq:per_ss}.

\section{Full counting statistics} \label{app:counting_statistics}

The stationary current and its noise can be derived  using full counting statistics as described in Refs.\,\cite{Schaller,Flindt2008,Emary2009}. Below, we give the main aspects of the derivation and the resulting expressions.

\subsection{Tilted Liouville operator}

The value of the particle current from lead $s$ integrated up to time $t$ equals the difference between the total numbers of electrons that have entered the dots from lead $s$ and that have  left the dots to that lead up to time $t$. For the parallel dots  initially in $\rho_\text{PD}(0)$, the characteristic function $\varphi_s(\chi,t)$ for its distribution is then encoded as
\begin{equation}
	\varphi_s(\chi,t)=\mathrm{Tr}\left\{e^{ t \mathcal{L}_{s}(\chi)}\left[\rho_\text{PD}(0)\right]\right\}
\end{equation}
by the tilted operator,
\begin{align}
   \mathcal{L}_s(\chi) =\mathcal{L}+ \sum_{\alpha=+,-}(e^{i\alpha\chi}-1)\mathcal{L}_{\alpha s}
\end{align}
where 
$\mathcal{L}_{\alpha s}(\rho_\text{PD})=J_{\alpha s}\rho_\text{PD}J_{\alpha s}^\dagger$
describe processes of exchanging  an electron between lead $s$ and the parallel dots, with $\alpha=+$ corresponding to the electron entering the dots, and $\alpha=-$ to the electron leaving.
 In particular, $ \mathcal{L}_s(\chi)$ reduces to the Liouvillian in \cref{eq:evolution2}   for $\chi=0$, so that $\varphi_s(0,t)=1$ as expected.

\subsection{Case of unique stationary state}

Since the $n$th cumulant of the integrated particle current equals the  $n$th derivative of the characteristic function, up to a factor $i^{n}$,  it is asymptotically linear in time when the stationary state $\rho_\text{PD}^\text{ss}$ is unique. In particular, the asymptotic rates for  average and the variance of the integrated particle current are given by [cf.~ \cref{eq:current}]
\begin{align}
	  \label{eq:Is}
   I_s =& \mathrm{Tr} \left[\left(\mathcal{L}_{+ s} - \mathcal{L}_{- s}\right)\left(\rho_\text{PD}^{\text{ss}}\right)\right]  ,\\
   \label{eq:Ss}
    S_s(0) =& \mathrm{Tr}\left[\left(\mathcal{L}_{+ s}  + \mathcal{L}_{- s}\right)\left(\rho_\text{PD}^{\text{ss}}\right)\right]
    \\
    &-2 \mathrm{Tr}\left[\left(\mathcal{L}_{+ s}  - \mathcal{L}_{- s} \right)\mathcal{R} \left(\mathcal{L}_{+ s}  - \mathcal{L}_{- s} \right)\left(\rho_\text{PD}^{\text{ss}}\right)\right],  \nonumber
\end{align}
respectively. Here, $\mathcal{R}(\rho_\text{PD})=\sum_{i\geq 2} \lambda_i^{-1} R_i \mathrm{Tr}(L_i \rho_\text{PD})$ is the reduced resolvent of $\mathcal{L}$ at $0$.

\subsection{Case of two stationary states}
In the case when the dynamics of \cref{eq:evolution3} features two stationary states denoted by $\rho_1^\text{ss}$ and $\rho_2^\text{ss}$ of \cref{eq:unp_stationary}, a general asymptotic state is their probabilistic mixture 
\begin{equation}	\label{eq:rhoss_deg}
\rho_\text{PD}^\text{ss}=p_1 \rho_1^\text{ss}+p_2\rho_2^\text{ss}.
\end{equation}
The probabilities are determined as  $p_1=\text{Tr}[ P_1\rho_\text{PD}(0)]$ and  $p_2=\text{Tr}[ P_2\rho_\text{PD}(0)]$ [cf.~\cref{eq:unp_proj}].
Then the average integrated current is also asymptotically linear in time, with the asymptotic rate as in \cref{eq:Is}, that is,
\begin{align}
	\label{eq:Is2}
	I_s =& p_1 I_{s1}+ p_2 I_{s2},
\end{align}
where $I_{s1}$ and $I_{s2}$ are the asymptotic rates for initial states found asymptotically in $\rho_1^\text{ss}$ and $\rho_2^\text{ss}$, respectively [or the average currents for those stationary states, cf.~\cref{eq:unp_current}].
In contrast, the variance of the integrated current in general diverges quadratically in time with the coefficient
\begin{align}
	\label{eq:Ss2}
	\sigma_s =& p_1 p_2 (I_{s1}- I_{s2})^2.
\end{align}
Only when the system is found asymptotically in either in $\rho_{1}^\text{ss}$ or $\rho_{2}^\text{ss}$, the variance is asymptotically linear in time, with the rates $S(0)_{s1}$ or $S(0)_{s2}$ given by \cref{eq:Ss}  with $\rho^\text{ss}_\text{PD}$ replaced by $\rho_1^\text{ss}$ or  $\rho_2^\text{ss}$. In fact, \cref{eq:Ss}  in general gives the rate of the asymptotically linear contribution to the variance with $\rho^\text{ss}_\text{PD}$ as in \cref{eq:rhoss_deg}.

\subsection{Case of perturbation away from two stationary states}

When the dynamics is perturbed away from the two-fold degeneracy of zero eigenvalue, Eqs.~\eqref{eq:Is} and \eqref{eq:Ss} can be expressed in the leading order as [Eqs.~\eqref{eq:Is} and \eqref{eq:Ss}]
\begin{align}
	\label{eq:Is_meta}
	I_s =& \ \mathrm{Tr} \left\{\left[\mathcal{L}_{+ s}^{(0)} \!-\! \mathcal{L}^{(0)}_{- s}\right]\!\!\left[\rho_\text{PD}^{\text{ss}(0)}\right]\right\} +...\\
	\nonumber=&
	\ p_1^\text{ss} I_{s1}+p_2^\text{ss} I_{s2}+...,\\
	\label{eq:Ss_meta}
	S_s(0) =& -\frac{2}{\lambda_2^{(2)}} \mathrm{Tr}\!\left\{\!\left[\mathcal{L}_{+ s}^{(0)} \! -\! \mathcal{L}_{- s}^{(0)} \right]\!\!\left[R_{2}^{(0)}\right]\! \right\}\\\nonumber&\qquad\qquad\qquad\times\mathrm{Tr} \!\left\{ L_2^{(0)}\!\! \left[\mathcal{L}_{+ s} ^{(0)} \!-\! \mathcal{L}_{- s}^{(0)} \right]\!\!\left[\rho_\text{PD}^{\text{ss}(0)}\right]\!\right\}+...\\
	\nonumber=&-\frac{2}{\lambda_2^{(2)}} p_1^\text{ss}p_2^\text{ss}\left(I_{s1}- I_{s2}\right)^2 +...,  
\end{align}
so that the fluctuation rate diverges inversely with the square of perturbation strength \cite{Macieszczak2021}, see also \cref{eq:per_ss,eq:per_decay}.

\begin{figure}[t!]
	\centering
	\includegraphics[width = 0.90\linewidth]{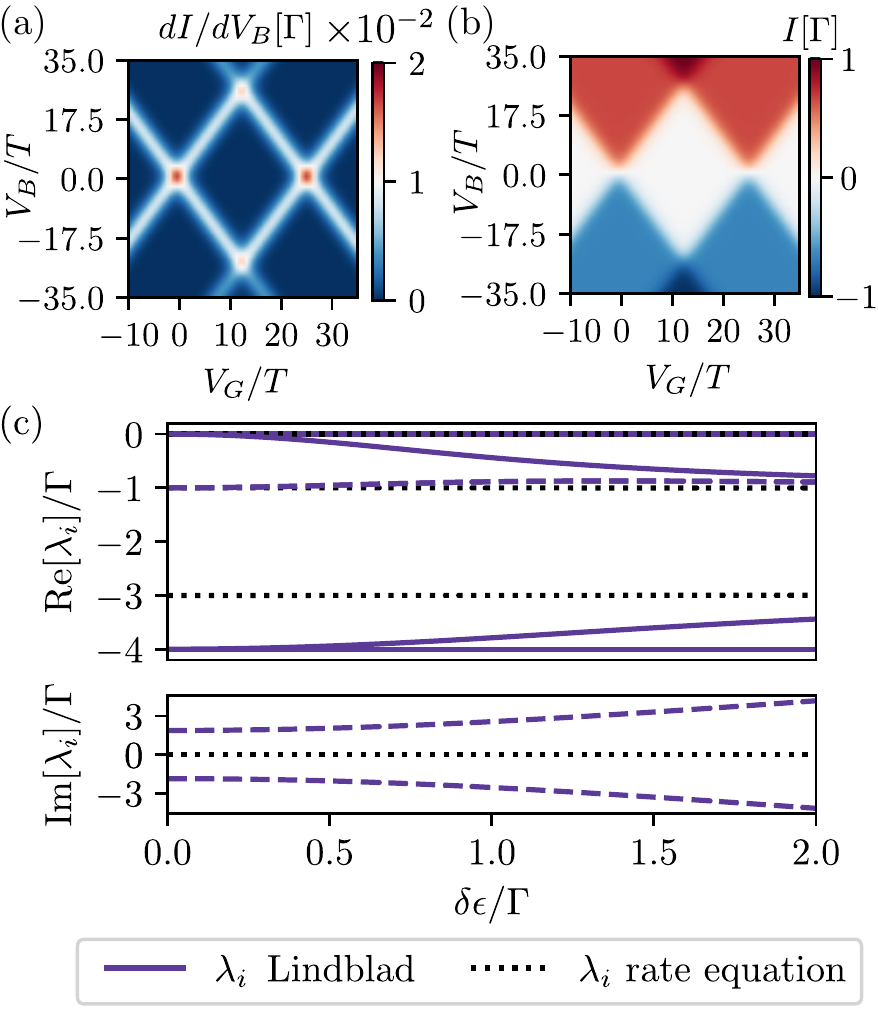}
	\caption{(a) Differential conductance and  (b) current of the parallel dots described by Pauli rate equation in the local basis.
		All other parameters are chosen as in \cref{fig:model}(b).
		(c) Spectrum of the Liouvillian [as in \cref{fig:spectrum}(a), purple solid lines indicating purely real $\lambda_1$, $\lambda_2$, $\lambda_5$, $\lambda_6$, while purple dashed lines the complex eigenvalues $\lambda_{3}$ and $\lambda_{4}$] and of the Pauli rate equation for the local basis (black dotted lines). For large $\delta \epsilon$, the coherences in the evolution are eliminated and the rate equation becomes exact. }
	\label{fig:diagonal_stability}
\end{figure}

\section{Pauli rate equation for parallel dots} \label{app:Pauli_stability}

Here, we compare the Pauli rate equation from \cref{eq:evolution} and consider the resulting stationary distributions. 
A Pauli rate equation for the diagonal entries of the density matrix in any basis can be obtained by neglecting the contribution from coherences.

In the basis $|00\rangle$,  $|10\rangle$, $|01\rangle$, and $|11\rangle$,  the Pauli rate equation \cite{Kirsanskas2017} features a single stationary probability distribution.  Even for the parameters chosen as in Eqs.~\eqref{eq:sym_par1} and~\eqref{eq:sym_par2}, which lead to stationary state degeneracy in the Lindblad dynamics of \cref{eq:evolution}, the distribution remains unique. 

\begin{figure}[h]
	\centering
	\includegraphics[width = 0.94\linewidth]{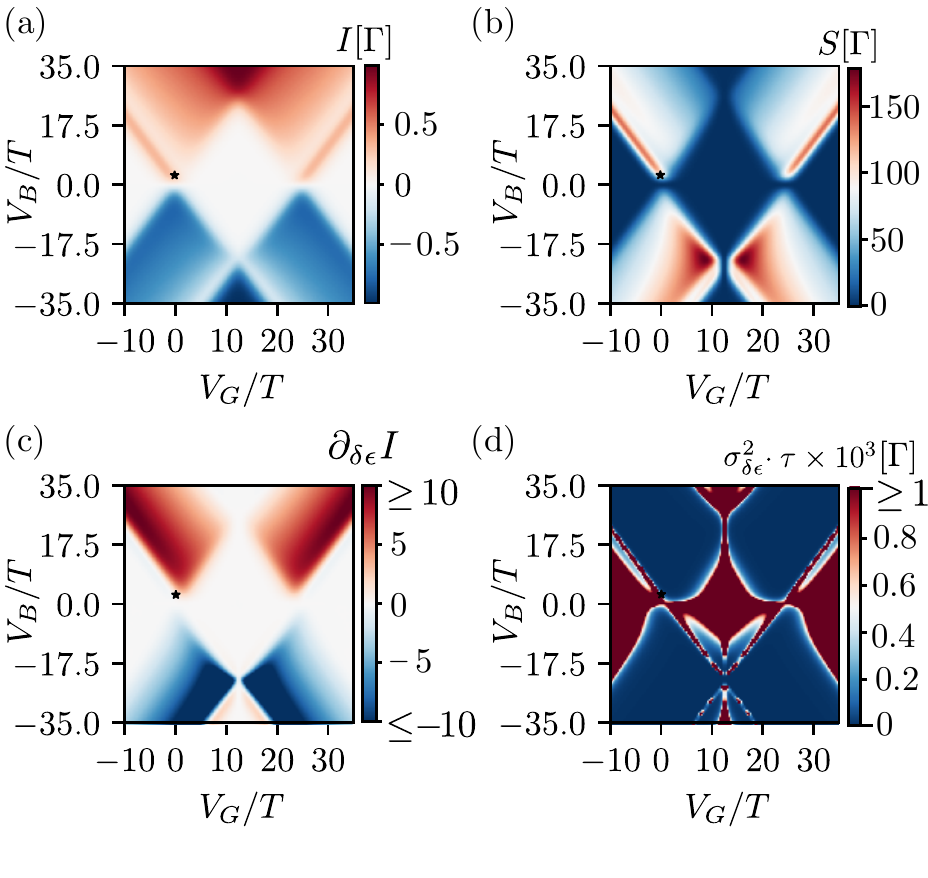}
	\caption{(a) Stationary  current, (b) noise, (c) signal rate and (d) the error with the same parameters as \cref{fig:model}(b) as functions of $V_B$ and $V_G$. The structure of the error in (d) within the Coulomb diamond is due to the current, noise and sensitivity not being exactly zero, but exponentially suppressed in this region. The star indicates the position in the stability diagram used for \cref{fig:sensor}.}
	\label{fig:stab_app}
\end{figure}

In \cref{fig:diagonal_stability}(a) and (b), the stationary differential conductance and stationary current for the rate equation are plotted \cite{Kirsanskas2017}.  The stability diagrams significantly differ from those for the stationary state of \cref{eq:evolution}, cf.~\cref{fig:model}(b) and \cref{fig:stab_app}(a). 
Here, the differential conductance recovers the typical Coulomb diamond structure for a single spinful dot.  
In fact, when $\epsilon_1=\epsilon_2=\epsilon$ and $\Gamma_{1s}=\Gamma_{2s}=\Gamma_{s}$, additionally with $\epsilon\ll U$, the dynamics corresponds to a single spin-degenerate dot, where coherences are suppressed because spin is a good quantum number in both leads and dot.  Such systems are used as charge sensors \cite{Bauerle2018}, where the parameters are chosen along the high conductance lines, where small changes in the gate voltage $V_G$ results in large response in the current.

For large detuning,  $| \epsilon_1-\epsilon_2| \rightarrow\infty $,  a Pauli rate equation captures the true evolution \cref{eq:evolution}. 
In the lowest order, classical dynamics arises between  $\lvert 00 \rangle\!\langle 00\rvert, \lvert 01\rangle\!\langle 10\rvert , \lvert 10\rangle\!\langle 10\rvert$, and $\lvert 11\rangle\!\langle 11\rvert$ that are left invariant by the Hamiltonian, and is given by the Pauli rate equation in the local basis.  The real parts of the eigenvalues for the rapidly oscillating coherences $|10\rangle\!\langle 01|$ and $|0 1\rangle\!\langle 10|$ are both given by $-\sum_{j=1,2}\sum_{s=L,R}\Gamma_{js}~[1-f_s(\epsilon_j)~+f_s(\epsilon_j+U)]/2$. 
The corresponding eigenvalue spectrum in \cref{fig:diagonal_stability}(c) indicate such behavior. 

\section{Stability diagrams} \label{app:stab_dia}

\cref{fig:stab_app} shows the non-trivial structure in the current and noise, and thus also in the sensitivity and error as functions of the gate and bias voltage.  The overall structure is caused by the Lamb shift and leads to an asymmetry in the bias voltage. It is therefore desirable to stay in the parameter regime that corresponds to lower noise when operating the parallel dots as a sensor and within this regime an optimal operation point may be found.

\end{document}